\documentstyle[preprint,aps,eqsecnum,epsfig]{revtex} 
\tighten 
\begin{document} 
\draft  
\title{\bf Dynamical viscosity of nucleating bubbles}
\author{\bf S. Alamoudi$^{(a)}$,D. G. Barci$^{(b)}$,
D. Boyanovsky$^{(a)}$,C. A. A. de Carvalho$^{(c)}$,\\
 E. S. Fraga$^{(c)}$, S. E. Jor\'as$^{(c)}$, F. I. Takakura$^{(d)}$ }
\address{$^{(a)}$Department of Physics and Astronomy, University of Pittsburgh,  
\\ Pittsburgh PA 15260, USA \\ 
$^{(b)}$Instituto de F\'\i sica, Universidade Estadual do Rio de Janeiro \\ 
Rio de Janeiro, RJ, 20559-900, Brasil \\ 
$^{(c)}$Instituto de F\'isica, Universidade Federal do Rio de Janeiro \\ 
C.P. 68528, Rio de Janeiro, RJ, 21945-970, Brasil \\ 
$^{(d)}$Departamento de F\'isica, Instituto de Ci\^encias Exatas \\
Universidade Federal de Juiz de Fora, Juiz de Fora, MG, 36036-330, Brasil}

\date{\today}
\maketitle 
\begin{abstract} 
We study the  viscosity corrections to the growth rate of 
 nucleating bubbles in a slightly supercooled first
order phase transition in $1+1$ and $3+1$ dimensional scalar field theory.  We propose a {\em microscopic} approach that leads to the non-equilibrium equation of motion
of the coordinate that describes small departures from the critical bubble and allows to extract the  growth rate consistently in a weak coupling expansion and in
the thin wall limit. Viscosity effects arise from the interaction of this coordinate with the stable quantum and
thermal fluctuations around a critical bubble.  In the case of $1+1$ dimensions we provide an estimate for the growth rate that depends on the
details of the free energy functional. In $3+1$ dimensions we recognize robust features that are a direct consequence of
the thin wall approximation that trascend a particular model. These are {\em long-wavelength hydrodynamic fluctuations}
that describe surface waves. We identify these low energy modes with quasi-Goldstone modes which are related to surface
waves on interfaces in phase ordered Ising-like systems. In the thin wall limit the coupling of this coordinate  to these hydrodynamic modes results in the largest contribution to the viscosity 
corrections to the growth rate. For a $\phi^4$ scalar field theory at temperature $T<T_c$ the growth rate  to lowest order in the quartic self-coupling $\lambda$ is $
\Omega = \frac{\sqrt{2}}{R_c}
\left[1- 0.003\lambda T \xi \left(\frac{R_c}{\xi}\right)^2 \right] $
with $R_c~;~\xi$ the critical radius and the width of the bubble wall respectively. We obtain the effective non-Markovian
Langevin equation for the radial coordinate and derive the generalized fluctuation dissipation relation, the noise is
correlated on time scales $\Omega^{-1}$ as a result of the coupling to the slow hydrodynamic modes. 
We discuss the applicability of our results to describe the growth rate of hadron bubbles in a  quark-hadron
 first order transition.

\end{abstract}

\newpage
\section{Introduction}

The dynamics of first order phase transitions is a fundamental ingredient in 
particle physics and in condensed matter. First order phase transitions occur
via the nucleation of  bubbles of the true vacuum phase in the metastable or false
vacuum phase. At large temperature it is mediated by thermal or overbarrier
activation and at low temperatures via quantum nucleation. First order phase transitions
are conjectured to occur in QCD  and in the Electroweak theory. In QCD a first order phase
transition {\em could} describe the hadronization of the quark-gluon plasma, possibly produced
in the early Universe at about $10^{-5}$ seconds after the Big Bang or in relativistic heavy 
ion collisions\cite{shuryak,meyer,kapu}. In the Electroweak theory a first order phase transition is argued to provide  the non-equilibrium setting for baryogenesis\cite{cohen,ruba,dolgov}. In early universe
cosmology, first order phase transitions had been proposed as mechanisms to generate the inflationary
stage\cite{linde,kolb,brand}. In condensed matter physics thermal activation results in the nucleation of bubbles
of the lowest free energy phase in binary 
fluids\cite{miguel,langercond} and also of the decay of metastable dimerized states in
quasi-one dimensional charge density wave systems and non-degenerate organic conductors\cite{yulu,boyancondbub}.  

The most comprehensive microscopic theory of nucleation via thermal activation was presented by Langer\cite{langer} and
was later extended to quantum field theory to account both for thermal activation as well as for quantum
nucleation\cite{coleman,stone,affleck,linde2}. An approach to describe nucleation in a non-steady state situation in
real time has been advocated in\cite{sphal}.  In the limit in which nucleation is dominated by thermal activation and
overbarrier transitions, these bubbles are produced via large thermal fluctuations. These bubbles will grow whenever their radius is larger than a critical value and collapse if
it is smaller. Supercritical bubbles will grow to convert the metastable phase into the stable phase or until they
percolate achieving the total conversion of the metastable phase. For slightly supercritical bubbles an important {\em dynamical} quantity is the growth rate of a 
bubble
\begin{equation}
\Omega = \frac{d}{dt}\left[\ln\left|\frac{R(t)}{R_c}-1\right|\right] ; ~~ R(t) \geq R_c 
\end{equation}
\noindent with $R(t)$ the (time dependent) radius of a slightly supercritical nucleating bubble and $R_c$ is
the critical radius\cite{langer,kapu,miguel}. Langer's theory provides the nucleation rate per unit volume per unit time
 given by
\begin{equation}
I= \Omega ~ {\cal D}~ e^{-\frac{F_{b}}{T}}
\end{equation}
\noindent where $F_b$ is the free energy of a critical bubble and ${\cal D}$ is proportional to the ratio of the determinants of the fluctuation operators around the bubble configuration to that around the homogeneous metastable state\cite{langer,kapu,miguel}. The regime of applicability of homogeneous nucleation theory is for $F_b >> T$, for
$F_b \approx T$ small amplitude thermal fluctuations can trigger the phase transition and nucleation and spinodal 
decomposition can no longer be distinguished. 
 The ratio of determinants in ${\cal D}$ 
has been computed analytically and numerically in several important cases\cite{maclerran,baacke,ramos}. Csernai and Kapusta\cite{kapu} have studied the growth rate of hadronic bubbles in a quark gas by extending
Langer's theory to the relativistic case. These authors studied a coarse-grained field theory in terms of a relativistic hydrodynamic description with viscous terms in the energy momentum tensor and the baryon current.
Their conclusion was that the growth rate is determined by the coefficients of the shear and bulk viscosities and that hadronic bubbles do not grow
when these coefficients vanish. Another approach presented by Ruggeri and Friedman\cite{ruggeri} also based on baryon-free relativistic hydrodynamics but with a different treatment of the heat conduction and energy flow  reached a different conclusion, th
at the growth rate is non-zero even for vanishing bulk and shear viscosities and that viscosity effects are subleading for small viscosities.
If the hadronization phase transition is of first order there are potentially important experimental signatures
associated with the homogeneous nucleation of the hadronic phase some of which had been studied in reference\cite{kapusta2}.

Homogeneous nucleation of the quark-gluon plasma has been recently studied with a bag model of
the equation of state for the quark and hadron phases and the different predictions above have been
compared\cite{zabrodin}.
The formulation and results of Csernai and Kapusta had recently been used to study first
order quark-hadron phase transition in the Early Universe for a first order hadronization transition\cite{chandra} and more recently homogeneous
nucleation has been tested numerically in $2+1$ dimensional systems with
 qualitative agreement to the standard result\cite{tetradis}.

An alternative phenomenological description of the growth rate based on
dissipative hydrodynamics combined with the finite temperature
effective potential has been presented in \cite{ignatius}. In this work analytical and numerical studies
 reveal a dependence of the growth rate on the phenomenological dissipative coefficient. 

Khlebnikov\cite{kleb} and Arnold\cite{arnold} have studied dissipative effects that slow down the growth rate of a supercritical bubble by coupling the order parameter  that describes nucleation to {\em other fields} (with a trilinear coupling) and applie
d the fluctuation dissipation relation to one loop order in the coupling to the other field. 

Finally we must mention a numerical approach to the description of  nucleation, based on a phenomenological
{\em Markovian} Langevin dynamics in terms of the finite temperature effective potential (typically computed to one
loop order) with a (local) friction coefficient and a Gaussian white noise term
related by the fluctuation-dissipation theorem\cite{gleiser,haas}. The use of the finite temperature effective potential in
the description of the inhomogeneous  bubble configuration that seeds the nucleation of the phase
of lowest free energy is an important ingredient, in particular in coarse-grained phenomenological descriptions\cite{kapu}. Finite temperature effects are included in the  coarse grained free energy which describes long wavelength physics by integrating o
ut short  wavelength fluctuations which are in thermodynamic equilibrium\cite{liu,kunz}. 
These finite temperature corrections modify quantitatively and sometimes
substantially the equilibrium free energy functional, for example the
position of the minima, masses and couplings. These are important parameters
that determine the profile of the bubble configuration, since the asymptotic
behavior of the bubble is determined by the position of the minima of
the equilibrium free energy. 

The importance of the growth rate for describing the dynamics of nucleation in the hadronization and
the Electroweak phase transitions  as well as the
practical importance of nucleation in quasi-one dimensional organic conductors justifies a continued effort to
understand from a {\em microscopic} perspective the influence of dissipation, viscosity and friction upon the
growth rate of supercritical nucleating bubbles.

{\bf Focus and Strategy:} Our goal in this article is  precisely to study dissipative effects upon the growth rate
 $\Omega$ from a more {\em microscopic} point of view in model quantum field theories that describe the main features
of nucleation. The  simplest model to study nucleation and a first order phase transition is a $\phi^4$ field theory with
a small explicit symmetry breaking term that breaks the degeneracy between ground states and thus leads to the existence of a metastable state. 

Although arguably this model could hardly describe the features of the quark-hadron or electroweak phase transitions,
our hope is to extract robust  phenomena that will be generic
to the physics of nucleation and that could enlighten the effect of viscosity on the growth rate.

The study begins by identifying the inhomogenous bubble configuration which is
a solution of the static equations of motion. We include the finite temperature effects by considering the equation
of motion in terms of the {\em finite temperature effective potential}, this is achieved consistently by the addition
of finite temperature counterterms to the Lagrangian. The quadratic fluctuations around this bubble  solution
feature an unstable mode that describes small perturbations from the critical bubble. The instability is a manifestation
of the growth (or collapse) of supercritical (or subcritical) bubbles and the unstable eigenvalue is directly related
to the growth rate of slightly supercritical bubbles. Besides this unstable mode there are modes of zero frequency 
corresponding to translations of this configuration and modes of positive frequency that correspond to stable fluctuations.

The main strategy is to consider the dynamics of the coordinate that determines the unstable direction in functional space, and treating it as a {\em collective coordinate}. We then focus on the {\em interaction} of this coordinate with those representing
 the stable fluctuations by expanding the original Lagrangian in terms of these coordinates in function
space and recognize the interaction terms between these coordinates.  The translational mode is anchored and ignored since we are not interested in studying the
dynamics of the translational degree of freedom but that of the growth of a supercritical bubble. We assume that the
stable degrees of freedom are in thermal equilibrium and that they serve as a ``bath'' with which the unstable coordinate
interacts. Integrating out
the degrees of freedom associated with the stable fluctuations we obtain an effective description of the dynamics for
the unstable coordinate which includes the ``viscosity and friction'' effects associated with the transfer of energy
to the stable modes. This description allows  us to obtain the equations of motion for the expectation value of the
unstable coordinate wherefrom we can obtain the {\em corrections} to the growth rate from the interaction with the
stable degrees of freedom. Furthermore, integrating out the stable degrees of freedom in the non-equilibrium functional
allows us to extract the effective Langevin equation for the unstable coordinate and to obtain consistently the noise
 terms.  We emphasize that this approach is {\em fundamentally different} 
from the previous studies described above. First of all, we are {\em not}
computing the ratio of determinants  that enter in
the rate expression. This ratio of determinants only involves the quadratic fluctuations but {\em not} the interaction between
the coordinates associated with the fluctuations. Our approach is 
different from that of references\cite{kleb,arnold}, in these references
the coupling of the order parameter associated with the nucleating field was to {\em other independent fields}. Contrary
to this we consider the coupling of the unstable coordinates to the stable fluctuations involving the {\em same}
field. Our approach starts from the microscopic theory and does not rely on a hydrodynamic description, however as it will be seen later, long-wavelength hydrodynamic fluctuations associated with surface waves provide the largest contribution to viscosity
 effects in three spatial dimensions.

In a very specific sense  our program obtains the effective action for the degree of freedom that
represents departures from a critical bubble by integrating out the degrees of freedom describing stable fluctuations, but
 of the {\em same field}. Our approach must necessarily rely on several different approximations: i) weak coupling
to allow a perturbative expansion of the effective action and the real time equations of motion, ii) the thin wall
approximation in which the critical radius is much larger than the width of the bubble wall. This approximation is
necessary to be able to provide quantitative answers. Finally as it will be justified later the important fluctuations
can be described by the {\em classical} limit of the finite temperature distribution functions. 

We carry out this program to lowest order in the quartic coupling  and to leading order in the thin wall approximation both in $1+1$ and $3+1$ dimensions. The motivation to study the lower dimensional case is provided
by its potential application in experimentally relevant condensed matter
systems\cite{yulu,boyancondbub}. 

{\bf Summary of results:} 

Our main results can be summarized as a consistent perturbative correction
to the growth rate from ``dynamical viscosity'' effects associated with
the interaction of the radius of the bubble with the stable quantum
and thermal fluctuations. In $1+1$ dimensions the growth rate for
bubbles is given by eqn. (\ref{estimate}) with $R_c,\xi$ the radius
and width of the critical bubble respectively, $\lambda$ the quartic
self-coupling and $F[R_c/\xi]$ a slowly
varying dimensionless function that depends on the details of the potential
and spectrum of stable fluctuations, in particular scattering states.  

In $3+1$ dimensions we find that in the thin wall limit the important
low energy excitations are {\em hydrodynamic fluctuations} associated with surface
waves. These are identified as {\em quasi-Goldstone} bosons and give the
dominant contribution to  viscosity effects on the growth rate. These low energy excitations are a robust feature of {\em any} model and not particular to the one considered in this article,
however, the {\em coupling} of the unstable radial coordinate to these fluctuations depends on the model.
  To lowest order in
the coupling the growth rate is given by eqn. (\ref{finalgrowthrate}). 

Furthermore, we obtain the effective Langevin equation for the unstable coordinate
that reveals the correct microscopic correlations of the noise term showing that the coupling to the
hydrodynamic modes determine that the noise is correlated on time scales $\Omega^{-1}$. 

The article is organized as follows: section II introduces the model,
describes the strategy in detail and sets up the general form of the
correction to the growth rate to lowest order (one loop) and in the
classical limit. In section III the case of $1+1$ dimensions is analyzed
in detail within the model considered. In section IV we study the $3+1$
dimensional case, and discuss the low energy fluctuations associated with
surface waves, providing the argument that these are quasi-Goldstone
hydrodynamic fluctuations. We argue that these low energy fluctuations
are present independent of the model and will dominate the viscosity
corrections to the growth rate. The conclusions are presented in section
V, along with a critical discussion on the validity of our results to describe the growth rate of supercritical bubbles in
coarse-grained descriptions of a first order quark-hadron phase transition. An appendix presents the effective Langevin equation for small departures of the critical radius and analyzes the generalized fluctuation-dissipation relation.

\section{The model: generalities and strategy}

We consider a real scalar field, $\phi(x)$, 
whose dynamics is determined by the following Lagrangian density
\begin{equation}
\label{lagrangeana}
{\cal L}={1\over 2}(\partial_\mu\phi)(\partial^\mu\phi)- V(\phi)\label{model},
\end{equation}
where  $V(\phi)$ is a double well potential with a metastable
minimum at $\phi_-$ and a stable minimum at $\phi_+$. It proves convenient to
parametrize it in the following form:

\begin{equation}
V(\phi)=\lambda (\phi-\phi_-)^2 \phi(\phi-\phi^*).
\label{pot}
\end{equation}

\begin{figure}
\centerline{ \epsfig{file=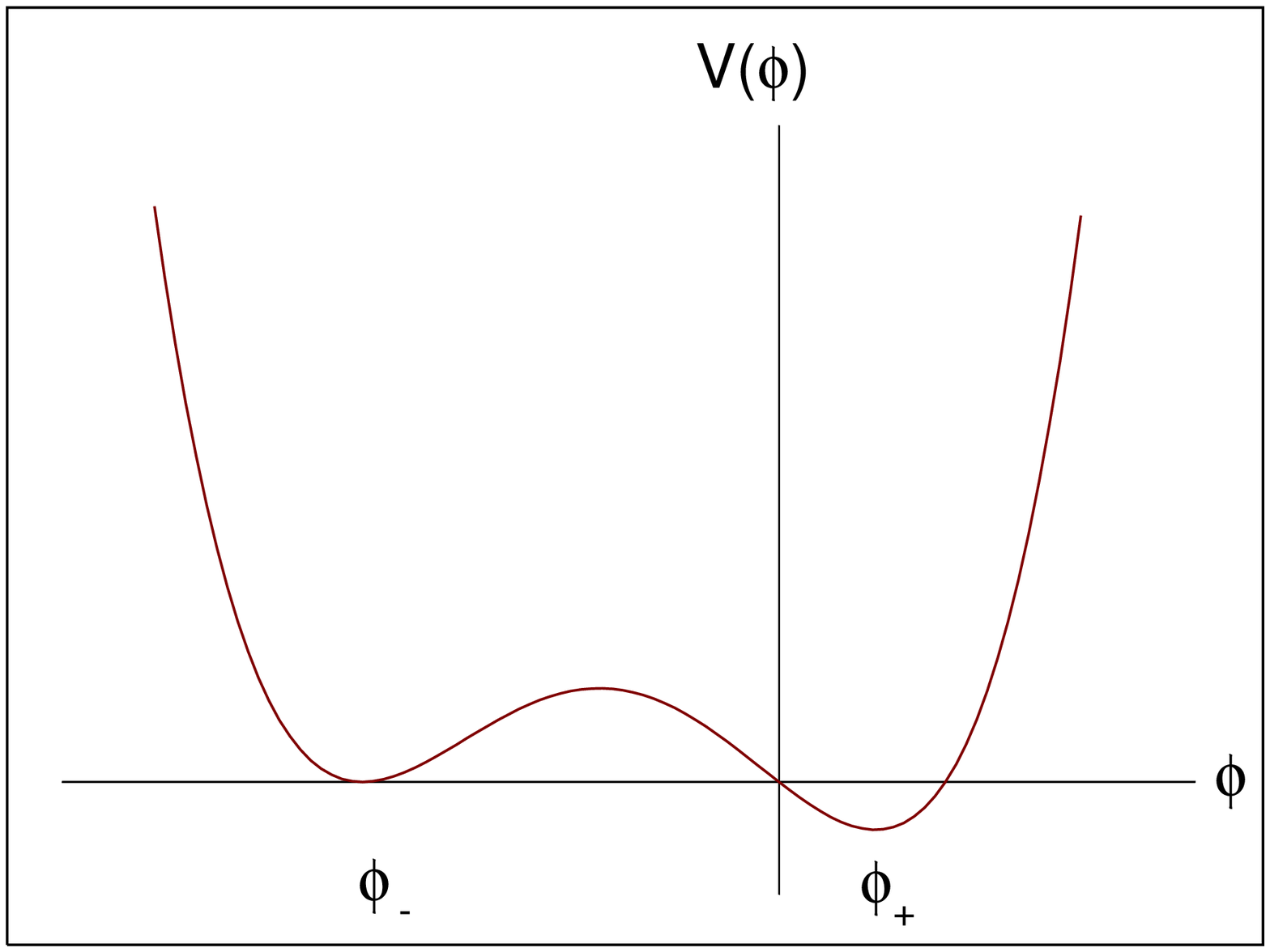,width=4.75in,height=3in}}
\caption{Form of the potential $V(\phi)$.\label{potfig}}
\end{figure}

This form of the potential is depicted in fig.(\ref{potfig}) and is very similar to the coarse-grained free energy
proposed in\cite{kapu} identifying the local energy variable {\em e} in that reference with the scalar field $\phi$, and
also similar to the effective potential description used in numerical simulations for the Electroweak theory\cite{gleiser,haas,kunz}.  

The stable minimum $\phi_+$ is given by
\begin{equation}
\phi_+ \,=\, \frac{1}{8} \left( 2 \phi_- + 3 \phi_* - \sqrt{4 \phi_-^2 + 9 \phi_*^2
-4 \phi_- \phi_*} \right).
\end{equation}
where the point $\phi_*$ is parametrized by the mass $m$ of small oscillations around the {\em metastable} 
minimum $\phi_-$ as
\begin{equation}
\phi^*\equiv\phi_-\left(1-\epsilon\right)
\end{equation}
where
\begin{equation}
\epsilon \equiv {m^2\over 2\lambda\phi_-^2} \;\;\;\; \mbox{and} \;\;\;\; m^2 \equiv 
V''(\phi_-)
\label{epsln}
\end{equation}
and $\lambda$ is the quartic coupling. Although this parametrization does not look familiar it
will prove advantageous later. In particular, $(1-\epsilon)$ gives a measure of the free energy difference 
 between the true and the false vacua and is therefore related to the supercooling temperature. As $\epsilon \rightarrow 1$, the two
minima become degenerate and the condition 
$\epsilon\approx 1$ defines the thin-wall limit (this will become clear later when we analyze the bubble profile), when the radius of the nucleating bubble is much larger than the width of its surface $\xi$ (or the correlation length). This
corresponds to the case of small supercooling $T<T_c$ but $T/T_c \approx 1$ in the description of reference\cite{kapu}.
In the thin wall limit, we find that $\phi_+$ has the following simple form
\begin{equation}
\phi_+ \,=\, -\,\frac{\phi_-}{2}(\epsilon-1) + {\cal O}((\epsilon-1)^2).
\label{phipls}
\end{equation}

\subsection{The counterterms:} 

We anticipate that there will be renormalization of the parameters in the potential, not only
arising from ultraviolet divergences but  more importantly from medium effects, i.e. finite 
temperature contributions to the mass, couplings and explicit symmetry breaking terms. In particular in three spatial dimensions we expect a correction to the mass  and the linear symmetry breaking term  both 
proportional to $T^2,T,\cdots$. The origin of these finite temperature corrections are the usual tadpole diagrams, the 
correction to the linear symmetry breaking term is a consequence of the trilinear coupling. These finite temperature
corrections are the usual ones obtained in an effective potential description and provide a finite renormalization to
the potential. Our aim is to describe the dynamical aspects of the thermal fluctuations that are responsible for thermal
activation and the nucleation of the true phase in the false vacuum phase. These are the solutions of the static
field equations but {\em in presence of the thermal fluctuations}, i.e. with the potential renormalized by the finite
temperature effects. That is to say, the potential that must enter in the field equations must be the finite temperature
effective potential\cite{liu,gleiser,kunz,tetradis,haas}. The finite temperature effective potential includes the finite temperature corrections for static and homogeneous field configurations. Including these finite temperature effects in the equations o
f motion guarantees that
 the {\em inhomogeneous} bubble configurations will asymptotically tend to the homogeneous values of the field representing the {\em correct} extrema of the equilibrium free energy. Furthermore, these finite temperature renormalizations will determine the
 {\em correct} scales for the bubble size, its width and the unstable frequency
associated with the growth of a supercritical bubble.  

We account for these  finite temperature renormalizations by replacing the parameters in
the potential by the finite temperature renormalized parameters and adding counterterms to the Lagrangian. The counterterms
are then found consistently in a coupling or loop expansion by requesting that they cancel the contributions from the
tadpole (or similar) terms {\em completely}, i.e. including the finite temperature parts. This prescription is at the
heart of using the finite temperature effective potential to study the solutions that lead to the decay of the metastable
state\cite{kapu,liu,gleiser,kunz,tetradis,haas}.  Thus the Lagrangian density becomes
\begin{equation}
{\cal L}[\phi] = {1\over 2}(\partial_\mu\phi)(\partial^\mu\phi)- V(\phi,T)+\delta {\cal L}_{ct}[\phi] \label{counter}
\end{equation}
where the potential $V(\phi,T)$ is of the same form as in (\ref{pot}) but in terms of the parameters that include
the finite temperature (and ultraviolet) renormalizations, i.e. $\lambda(T), m(T),\epsilon(T)$ etc. and the
counterterm Lagrangian density is of the form
\begin{equation}
\delta {\cal L}_{ct} = \delta \lambda \phi^4 + \delta g \phi^3+ \delta M^2 \phi^2 + H\phi   \label{countertermlagra}
\end{equation}
where space-time translational invariance dictates that the 
counterterms are constant.
These counterterms will be consistently chosen in the loop expansion to cancel the contribution from local tadpoles.
We will argue below that this procedure renormalizes consistently in
perturbation theory the {\em static} properties of the bubble, in particular the width and radius of a critical bubble.
 In particular the free energy density studied in\cite{kapu,gleiser,haas,kunz,liu} is precisely of this form with
parameters that depend on temperature. 

At this point we may consider adding to the counterterms a wave-function renormalization which 
would involve corrections for inhomogeneous configurations. A wave-function renormalization will 
arise at one loop order and beyond because
of the trilinear coupling. However, computing this correction in the usual loop expansion around a homogeneous background
is not very relevant for the bubble configuration. Since our goal is to obtain the {\em real time} equations of motion
for particular fluctuations around the bubble configuration, these renormalizations will arise automatically in the
equations of motion obtained to one loop order. Thus the counterterms that we include in the Lagrangian density are only
those for {\em static} renormalization thus accounting for the correct scales and asymptotic values of the classical
solution and obtain the equivalent of the wave-function renormalization directly from the
real time evolution of fluctuations around the bubble configuration. This procedure will become clear when we 
obtain explicitly 
the evolution equations in the next sections. 

We will focus our attention on the cases of one and three spatial dimensions. Although the case of one spatial
dimension is of limited interest in particle physics, it is important in condensed matter and statistical physics of
quasi-one dimensional systems\cite{yulu,boyancondbub}. Furthermore most aspects of the treatment are general and the one-dimensional case provides a somewhat simpler setting to introduce the main strategy as well as to explore the general features. 
The three dimensional case will be studied  in detail subsequently.

\subsection{The bubbles:} 
Before proceeding to the specific situations, we now consider general features of the bubble solutions and
the quadratic fluctuations around them to focus on the precise strategy to follow. 

A critical bubble is a solution of the {\em static} field equations, which  in terms
of the (effective) potential $V(\phi,T)$ is given in one spatial dimension by
\begin{equation}
\left. -\frac{d^2\phi_b}{dx^2}+\frac{\partial V(\phi,T)}{\partial\phi}\right|_{\phi_b}\,=\,0 \label{1Dbubble}
\end{equation}

\noindent with boundary conditions such that 
$ \phi_{b} \rightarrow \phi_- $ for $\vert x \vert \rightarrow \infty $. Such a solution describes a ``bubble'' like configuration which
approaches the false vacuum at asymptotically large distances and probes the true vacuum in a localized region in
space of size $2R_c$ with $R_c$ the critical ``radius''.  Such a solution  will be found in detail in  section III. In three spatial dimensions, the critical bubble is a radially symmetric
static field configuration solution of the following  equation 

\begin{equation}
\left. - \frac{d^2 \phi_{b}}{dr^2} - \frac{2}{r} \frac{d \phi_{b}}{dr} + \frac{
\partial V(\phi,T)}{\partial\phi}\right|_{\phi_b}\,=\,0
\label{3Dbubble}
\end{equation}

\noindent with the boundary condition 
$\phi_{b}(r) \rightarrow \phi_-$ as $r \rightarrow \infty$. 
It corresponds to a field 
configuration that starts close  to the true vacuum, $\phi_+$, and tends to
the false vacuum, $\phi_-$, at asymptotically large radial distance. The
solution of (\ref{3Dbubble}) will be studied in detail in the thin wall
approximation in section IV.

In both cases the change from the true vacuum to the 
false one occurs around the radius of the bubble, $R_c$, over a distance $\xi(T) \sim 1/m$ that defines
 the wall thickness of the bubble and is related to the correlation length in the metastable phase.
 We will study the non-equilibrium dynamical
viscosity of nucleating bubbles in the thin wall limit in which the radius of the
bubble is much larger than the wall thickness; i.e. $R \gg \xi$. These field configurations are parametrized
by two important coordinates: $\vec{x}_0$ which describes  the position of the center of the bubble and
the radius $R$, i.e. a bubble configuration is of the form $\phi(\vec x - \vec{x}_0;R)$ and a critical bubble
corresponds to $R=R_c$ determined by the solution to the equations of motion (\ref{1Dbubble},\ref{3Dbubble}).
 The coordinate ${\vec x}_0$ is associated with translational invariance and typically
treated as a collective coordinate\cite{rajaraman}, while the radius $R$ determines the size of the bubble and will
be treated also as a collective coordinate (see below). Since we are not concerned here with the dynamics of the
translational degrees of freedom, but rather with that of the radius, we ``clamp'' the collective 
coordinate $\vec{x}_0$ by fixing the center of the bubble at $\vec{x}_0=0$. Integration over this collective coordinate results in the typical volume factor\cite{langer,coleman} and is not relevant for our discussion.

In $d$ space dimensions, we can use the radius of the bubble $R$ as a variational parameter and introduce the total energy of a bubble configuration with radius $R$  by
\begin{equation}
E_{var}(R) = \int d^d x\;\left[\frac{1}{2}\left(\nabla \phi_{b}(\vec x,R) \right)^2 + 
V(\phi_{b}(\vec x, R))\right].
\label{bubenergy}
\end{equation}
and in the limit $R/\xi(T)>>1$ (thin wall) has two main contributions: a  volume contribution proportional to $R^d$
which is negative and arises from the region of the bubble that probes the true vacuum which has lower (free) energy, and
a surface term which is positive and arises from the region of the bubble corresponding to the wall,
 includes gradient terms and is
therefore proportional to the area of the surface of the wall, $R^{d-1}$. In one spatial dimension the ``surface'' of
the bubble corresponds to two points and the gradient energy saturates to a constant independent of the
radius for large bubble radius.

\begin{figure}
\centerline{ \epsfig{file=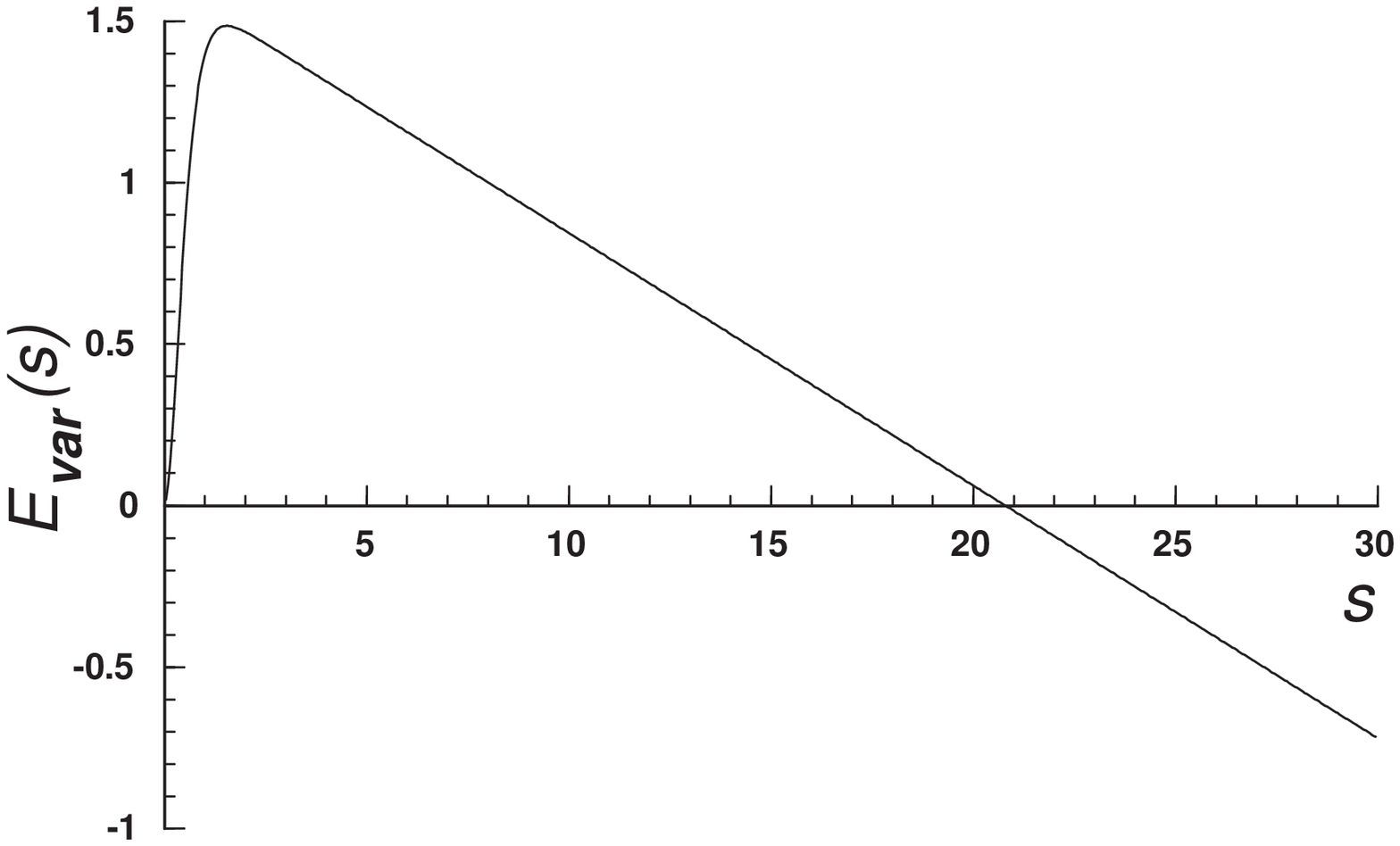,width=4.75in,height=3in}}
\caption{The total energy of the one dimensional bubble as a function of its dimensionless radius $s$ for $\epsilon = 1.2, \lambda = 0.1$ and $\phi_- = 2.0$. \label{e1dfig}}
\end{figure}

\begin{figure}
\centerline{ \epsfig{file=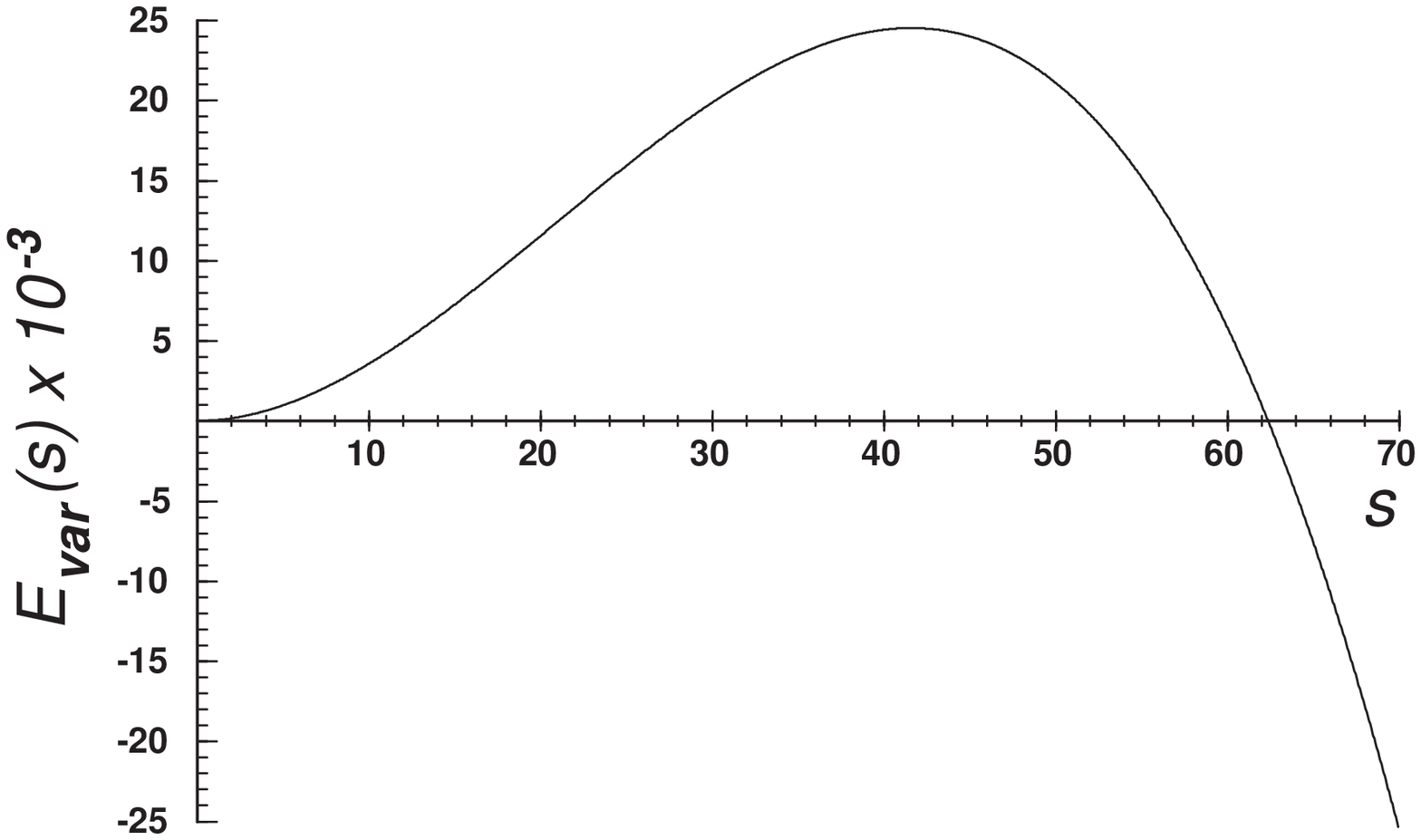,width=4.75in,height=3in}}
\caption{The total energy of the three dimensional bubble as a function of its dimensionless 
radius $s$ for $\epsilon = 1.2, \lambda = 0.1$ and $\phi_- = 2.0$. \label{e3dfig}}
\end{figure}

The general behavior of the total energy 
of a bubble configuration as a function of  $s=R/\xi$ is depicted in fig. (\ref{e1dfig}) for one spatial dimension  and in fig. (\ref{e3dfig}) for three spatial dimensions (see sections III and IV). The maximum of the energy function determines the value 
of the critical radius $R_c$ as discussed in detail in the next sections. 
Clearly a critical bubble is an {\em unstable} static solution of the equations of motion. For $R<R_c$ the surface energy
term dominates and the bubble shrinks into the false vacuum phase, for $R>R_c$ the volume energy dominates and the bubble
 will grow as the gain in the volume energy is greater than the cost in elastic surface energy.

A more clear description is obtained in the case of three spatial dimensions for a spherically symmetric bubble. 
Let $\sigma$ be the surface tension (energy per unit area) of the spherical 
bubble configuration separating
the metastable phase from the stable one. Then the total energy of the bubble is 
\begin{equation}
E_{var}(R) = -\frac{4 \pi}{3} \Delta {\cal F} R^3 + 4 \pi \sigma R^2 + \cdots
\label{classengy}
\end{equation}
where $\Delta {\cal F}= \left| V(\phi_+,T) - V(\phi_-,T) \right|$, and the dots stand for  corrections that are subleading in the thin wall
approximation, see section IV (eqn. (\ref{energy3dbubb}) for details. These (small) corrections will be neglected for the arguments presented
below.  
The energy attains its maximum $E_c$ at the critical radius  
\begin{equation}
R_c = \frac{2 \sigma}{\Delta {\cal F}}
\label{rc}
\end{equation}
 and it is given by
\begin{equation}
E_c \equiv E_{var}(R_c)=  \frac{4 \pi \sigma}{3} R_c^2 \,= \, \frac{16 \pi \sigma^3}{3 \left(\Delta {\cal F}
\right)^2}.
\label{ec}
\end{equation}
Near the maximum of the energy function, it can be written as an expansion in terms of the (small) departures
from the critical radius as
\begin{equation}
E_{var}(R) = E_c - \frac{1}{2}\omega^2 (R-R_c)^2 + \cdots \label{energyofR}
\end{equation}
where the frequency
\begin{equation}
  -\omega^2 \equiv \left. \frac{d^2 E_{var}(R)}{dR^2} \right|_{R_c}  = - 8 \pi \sigma
\label{wsecdv}
\end{equation}
is independent of the critical radius and $\Delta {\cal F}$ and as it will
be seen in detail below, it is related to the
growth rate of slightly supercritical bubbles.

\subsection{Fluctuations}

The study of the fluctuations around the classical bubble configuration begins by studying the spectrum of
the fluctuation operator
\begin{eqnarray}
&& \left[-\frac{d^2}{dx^2}+V''(\phi_b(x,R_c))\right] {\cal U}_n(x) = \omega^2_n {\cal U}_n(x) ~~\mbox{in 1D} \label{1dfluc}\\
&& \left[-\vec{\nabla}^2+V''(\phi_b(\vec x,R_c))\right] {\cal U}_n(\vec x) = \omega^2_n {\cal U}_n(\vec x) ~~\mbox{in 3D} 
\label{3dfluc}
\end{eqnarray}
where the prime represents differentiation with respect to the field $\phi$.

Taking a spatial derivative of the equation of motion satisfied by the bubble configuration it is straightforward to
find that
\begin{equation}
{\cal U}_0(x,R_c) \propto \frac{d\phi_b(x,R_c)}{dx} \label{1dzeromode}
\end{equation}
is a solution of the one dimensional eigenvalue problem (\ref{1dfluc}) with $\omega^2_0=0$ and that
\begin{equation}
{\cal U}_0(\vec x,R_c) \propto \frac{\vec x}{|\vec x|}\frac{d\phi_b(r,R_c)}{dr} \label{3dzeromode}
\end{equation}
are eigenfunctions of the three dimensional fluctuation operator with zero eigenvalue. These are the zero modes
arising from translational invariance\cite{langer,coleman,rajaraman}.  In one dimension the zero mode is an odd function of $x$ 
since the bubble solution is even, therefore it has a node. Hence there must be another solution of the Schroedinger
like operator (\ref{1dfluc}) that has no nodes and has a smaller  and therefore {\em negative} eigenvalue. The
form of the bubble solution in the thin wall approximation is that of a widely separated kink-antikink pair, as it will be shown in detail in the next section we find that the eigenfunction corresponding to the negative
eigenvalue must be  ${\cal U}_{-1} \propto d\phi_b(x,R)/dR |_{R_c}$. This can be understood simply
from the fact that the zero mode associated with the bubble solution is the {\em antisymmetric} combination of the
zero modes associated with the individual kinks\cite{rajaraman}, hence the eigenfunction with lower eigenvalue must be
the {\em symmetric} combination and the form of the bubble profile immediately leads to the conclusion that is the
derivative with respect to the radius (see  section III, eqn. (\ref{1dunstmode})). In the three dimensional case, the zero modes (\ref{3dzeromode}) correspond to
the angular momentum $l=1$ spherical harmonics, therefore there must be an $l=0$ (spherically symmetric) solution of
the Schroedinger-like eigenvalue problem (\ref{3dfluc}) with a {\em negative eigenvalue}. In a later section we study in detail the bubble solution and
the spectrum of fluctuations around it and  conclude that the eigenfunction 
corresponding to the negative eigenvalue is ${\cal U}_{-1} \propto d\phi_b(r,R)/dR|_{R_c}$ in the thin wall approximation 
(see section IV, eqn. (\ref{spcquasi}) for ${\cal U}_{000}$). 

Therefore the respective spectra of the fluctuation operators are
\begin{eqnarray}
&& {\cal U}_{-1}(\vec x,R_c) = \sqrt{N_{-1}}~ \left.\frac{d\phi(\vec x,R)}{dR}\right|_{R_c}, 
~{\omega^2_{-1} = -\Omega^2 \; ; \; \Omega^2 >0 } \; \; ; \; \; {\cal U}_{0}(\vec x,R_c) = \sqrt{N_0}~ \vec{\nabla}\phi(\vec x,R_c), ~{\omega^2_0 = 0} \nonumber \\ 
&& {\cal U}_{n>0}(\vec x,R_c), 
~{\omega^2_n > 0} \label{spectrum}
\end{eqnarray} 
with $N_{-1}; N_0$ normalization factors (chosen real). 
The classical bubble solution corresponds to a saddle point in functional space, the mode ${\cal U}_{-1}$ determines
an unstable direction. 

The field operator can now be expanded in the complete basis of eigenfunctions of (\ref{1dfluc},\ref{3dfluc}) in either
case and write in general

\begin{equation}
\phi(\vec x,t)=\phi_b(\vec x-\vec{x}_0,R_c)+ q_{-1}(t){\cal U}_{-1}(\vec x-\vec{x}_0)+ q_0(t){\cal U}_{0}(\vec x-\vec{x}_0)+\sum_{n>0} q_n(t){\cal U}_n(\vec x-\vec{x}_0), \label{fieldexpansion}
\end{equation}
Whereas the bound state eigenfunctions are chosen real, the scattering states are chosen to satisfy the
hermiticity condition ${\cal U}_{n}(\vec x) = {\cal U}^*_{-n}(\vec x); ~~q_n(t)= q^*_{-n}(t)$.

The quanta associated with the coordinates $q_{n>0}$ for  the stable modes will be referred generically as ``mesons'', 
and the operators $q_{n>0}$ create and annihilate meson states.

From (\ref{spectrum}) it is clear that the modes ${\cal U}_0 \; ; \; {\cal U}_{-1}$ correspond to  shifts in the
position of the bubble, $\vec{x}_0$ and the radius $R$ respectively.  The collective coordinate treatment\cite{rajaraman} absorbs the zero mode ${\cal U}_0$ into a definition of a new quantum mechanical degree of freedom $\hat{\vec{x}}_0(t)$ and
expands the field operator in terms of the directions perpendicular to this zero mode,
\begin{equation}
\phi(\vec x,t)=\phi_b(\vec x-\hat{\vec{x}}_0(t),R_c)+ q_{-1}(t){\cal U}_{-1}(\vec x-\hat{\vec{x}}_0(t))+ 
\sum_{n>0} q_n(t){\cal U}_n(\vec x-\hat{\vec{x}}_0(t))\label{expre}
\end{equation}

We are not interested in the dynamics of the translational collective coordinate, therefore we will ``clamp'' the
position of the bubble and focus solely on the dynamics of the unstable coordinate $q_{-1}(t)$. 
Technically this is achieved by inserting a functional delta function for $\hat{\vec{x}}_0(t)$ in 
the path integral, 
the corresponding Jacobian from the
change of field variables to the collective coordinate\cite{rajaraman} leads 
to the usual volume factor\cite{langer,coleman}, hence in what follows we will set 
$\hat{\vec{x}}_0(t)=0$. 
We note that for small amplitudes of the unstable mode we can write the
expansion (\ref{expre}) as
\begin{equation}
\phi(\vec x,t)\approx \phi_b(\vec x,R(t))+ 
\sum_{n>0} q_n(t){\cal U}_n(\vec x) \; ; \; \; 
R(t)=R_c+q_{-1}(t)\sqrt{N_{-1}}\label{expre2}
\end{equation}
that allows us to identify  $R(t)-R_c=q_{-1}(t)\sqrt{N_{-1}}$ and
the coordinate $q_{-1}$ is associated with  small departures
from the critical radius.

\subsection{The relation between $\Omega^2$ and $\omega^2$}

At this stage we recognize that the fluctuation mode ${\cal U}_{-1}(\vec x)\propto d\phi_b(\vec x,R_c)/ dR_c$ is the (unstable) functional direction along which the bubble either grows or collapses, therefore the coordinate $q_{-1}(t) \propto (R(t)-R_c)$ 
describes the small fluctuations around the critical bubble. An immediate question is: what is
the relation between the frequency $\omega^2$ in (\ref{energyofR},\ref{wsecdv}) and $\Omega^2$, the eigenvalue 
associated with ${\cal U}_{-1}$ (\ref{spectrum})? The answer to this important question is obtained by 
taking the second derivative of the total energy in eq. (\ref{bubenergy}) with respect to
the radius of the bubble $R$, which results in the following equation
\begin{equation}
\frac{d^2 E_{var}(R)}{d R^2} \,=\, \int d^dr \, \left\{ 
\left[\vec{\nabla}\left(\frac{d\phi_b}{dR}\right)\right]^2+
\vec{\nabla}\phi_b.\vec{\nabla}\left(\frac{d^2\phi_b}{dR^2}\right) +
\left(\frac{d^2\phi_b}{dR^2}\right) \frac{\partial V(\phi_b)}{\partial \phi} +
\left(\frac{d\phi_b}{dR}\right)^2 \frac{\partial^2 V(\phi_b)}{\partial \phi^2}\right\}
\label{d2ER}
\end{equation}

Integrating the first two terms in eq. (\ref{d2ER}) by parts using the boundary conditions for the bubble
solution and eqs. (\ref{1Dbubble},\ref{3Dbubble},\ref{1dfluc},\ref{3dfluc}), which are evaluated at the critical radius $R_c$, we obtain
the following relation
\begin{equation}
\left.\frac{d^2 E_{var}(R)}{d R^2}\right|_{R=R_c} \,=\, - \omega^2 \,=\,
-\frac{\Omega^2}{N_{-1}}
\label{omegarelation}
\end{equation}
in terms of the normalization factor of the unstable mode (see (\ref{spectrum})) which will be found for the one
and three dimensional cases separately below. We note that the relation
(\ref{expre2}) determines that in the absence of interactions with other
degrees of freedom the growth rate of a supercritical bubble
is given by
\begin{equation}
\frac{d}{dt}\left[\ln\left|\frac{R(t)}{R_c}-1\right|\right]= \Omega \label{growthrate}
\end{equation}
for small departures from the critical radius. 

Equation (\ref{omegarelation}) is very useful since it relates the second derivative of
the total {\em classical} variational energy $E_{var}(R)$ (the energy of the classical bubble as a function of its
radius) to the eigenfrequency associated with the {\em quantum} unstable mode. Usually
it is a difficult task to find the spectrum of the quantum fluctuation operator, and
consequently the frequency of the unstable mode. The above relation can be used 
to find the value of the unstable frequency, or the normalization factor
$N_{-1}$ if the frequencies are known.

\subsection{The strategy}
Our goal is to study the effects of quantum and thermal fluctuations upon the {\em real time} evolution
of a bubble whose radius is slightly larger than critical. This is achieved by treating the coordinate $q_{-1}(t)\propto
(R(t)-R_c)$ as a {\em collective coordinate}, i.e. the radius of the bubble is  treated as a fully quantum mechanical variable interacting with the other degrees of freedom corresponding to the stable fluctuations and described by the
coordinates $q_n(t)$. The ``zero mode'' (translational degree of freedom) is clamped and frozen since we are only interested
in describing the dynamics of the unstable mode. The interaction of the stable degrees of freedom with the collective
coordinate describing the departure from the critical radius will introduce viscosity effects and slow down the growth
of a supercritical bubble. This will result in a smaller growth rate and our aim is to precisely compute this viscosity
effect on the growth rate for small departures from the critical radius. By including the finite temperature {\em static}
counterterms in the Lagrangian and requesting that these be cancelled consistently in the perturbative expansion, we
can isolate the static renormalization to the unstable frequency from the {\em dynamical} viscosity effects associated
with the energy transfer to the stable modes.  

This program begins by expanding the field in terms of the unstable and stable coordinates (after clamping the translational
mode) as

\begin{equation}
\phi(x,t)=\phi_b(x,R_c) + q_{-1}(t) {\cal U}_{-1}(x)+
\sum_{p} q_p(t) {\cal U}_p(x),
\label{expansion}
\end{equation}
where the summation index $p$ runs over all stable, bound and scattering, states other than
the translational and unstable  modes. Throughout the index $p$  refers to both scattering and bound states.

Using the above expansion, eq. (\ref{expansion}), of the field $\phi(x,t)$, the Lagrangian can be shown to have the following form:
\begin{equation}
L[q_{-1},q_p]\,=\,L_0[q_{-1}] + L_0[q_p] + L_I[q_{-1},q_p]
\end{equation}
where
\begin{eqnarray}
L_0[q_{-1}]&=&\frac{1}{2}~\left\{ \dot q_{-1}^2(t)+
\Omega^2 q_{-1}^2(t)+ \delta \Omega^2 q_{-1}^2(t) + hq_{-1}(t)\right\}  \label{unlag} \\
L_0[q_{p}] &=& \frac{1}{2}~\sum_{p}~
\left\{ \dot{q}_p(t)\dot{q}_{-p}(t)-\omega_p^2 q_p(t)q_{-p}(t)\right\} \label{scatlag} \\
L_I[q_{-1},q_p] &=& - \sqrt{\lambda}~q_{-1}(t)
\sum_{p,p'}{\cal B}_{pp'}q_p(t)q_{p'}(t)
\, - \, q_{-1}^2(t)
\left\{ \sqrt{\lambda} \sum_{p}{\cal B}_pq_p(t) + \lambda\sum_{p,p'}{\cal A}_{pp'}
q_p(t)q_{p'}(t)\right\} \nonumber \\
&& -\; q_{-1}^3(t) \,\left\{\sqrt{\lambda}\,{\cal B}_{-1} \,+\,\lambda\, \sum_{p}
{\cal A}_pq_p(t) \right\} -\,\lambda \, {\cal A}_{-1}\,q_{-1}^4(t)+\mbox{h.o.t}+
L_{ct}[q_{-1},q_p], \label{lagrainte}
\end{eqnarray}
where we have used the equations of motion satisfied by the bubble
configurations,  and equations (\ref{1dfluc}) and (\ref{3dfluc}). The
terms $\delta \Omega^2 q^2_{-1} \; ; \; h q_{-1}$ are the quadratic and
linear terms in $q_{-1}$ from the counterterm Lagrangian. The terms denoted
by h.o.t. in (\ref{lagrainte}) correspond to higher order interactions
that are not important to the order that we are studying. 
The term $L_{ct}[q_{-1},q_p]$ arise from the non-linear (cubic and higher)
terms in the coordinates  from the counterterm Lagrangian, they do not need
to be specified since they will be requested to cancel tadpole and static
terms in the equations of motion for the collective coordinate $q_{-1}$ and do not 
contribute to lowest order (${\cal O}(\lambda)$). 
The matrix elements are given by
\begin{eqnarray}
{\cal B}_{pp'} & \equiv & \frac{1}{2\sqrt{\lambda}}\int_{-\infty}^{+\infty}~dx'~
{\cal U}_{-1}(x'){\cal U}_p(x'){\cal U}_{p'}(x')V'''(\phi_b(x',R_c)),
\label{Bpp}\\
& & \nonumber \\
{\cal B}_p & \equiv & \frac{1}{2\sqrt{\lambda}}\int_{-\infty}^{+\infty}~dx'~
{\cal U}_{-1}^2(x'){\cal U}_p(x')V'''(\phi_b(x',R_c)),
\label{Bp}\\
& & \nonumber \\
{\cal B}_{-1} & \equiv & \frac{1}{6\sqrt{\lambda}}\int_{-\infty}^{+\infty}~dx'~
{\cal U}_{-1}^3(x')V'''(\phi_b(x',R_c)),
\label{B}\\
& & \nonumber \\
{\cal A}_{pp'} & \equiv & 6\int_{-\infty}^{+\infty}~dx'~
{\cal U}_{-1}^2(x'){\cal U}_p(x'){\cal U}_{p'}(x'),
\label{Akk} \\
{\cal A}_{p} & \equiv & 4\int_{-\infty}^{+\infty}~dx'~
{\cal U}_{-1}^3(x'){\cal U}_p(x'),
\label{Ak} \\
& & \nonumber \\
{\cal A}_{-1} & \equiv & \int_{-\infty}^{+\infty}~dx'~
{\cal U}_{-1}^4(x'),
\label{A}
\end{eqnarray}
so that the $\lambda$-dependence is explicit in each term of the 
effective action.

We thus see that the above Lagrangian describes a quantum mechanical
degree of freedom $q_{-1}$ interacting with a bath of infinitely many
degrees of freedom $q_p$, as well as with self-interaction. The 
self interaction of the coordinate $q_{-1}$ will be neglected because 
we are only interested in the viscosity effects arising from interaction
with the stable degrees of freedom. The non-linear terms in $q_{-1}$ are
neglegted because we will extract the growth rate for small departures from the critical bubble. 

The bath of mesons will introduce viscosity effects on evolution of
the coordinate associated with departures from the critical radius and
will result in a correction to the growth rate. These effects can be 
studied in a consistent manner by integrating out the meson degrees of freedom in a consistent perturbative expansion, thus obtaining the
non-equilibrium effective action for $q_{-1}$. The variational derivative 
of this effective action leads to the equation of motion that includes the effects of the meson bath. This non-equilibrium effective action is
obtained up to one loop order in appendix B. The effective equation of
motion is a Langevin equation with a non-Markovian viscosity kernel and
a Gaussian noise term (to one loop order) whose correlations are related to the viscosity kernel via the Fluctuation-Dissipation relation. The details are provided in appendix B. 

 Alternatively we obtain the equation
of motion for the expectation value of the coordinate $q_{-1}$ directly by means of linear response. 

 We have already 
accounted for the {\em static} renormalization of the unstable
frequency $\Omega$ by introducing the finite temperature counterterms
and using the finite temperature effective potential in the equations
of motion. However, we anticipate that there will arise further corrections
to the growth rate $\Omega$ from {\em velocity dependent} terms in the
effective equations of motion. These are truly {\em dynamical} viscosity
effects which cannot be captured by a static calculation. The viscosity
terms that are a function of the time derivative of the unstable coordinate
will be revealed by obtaining the effective non-equilibrium equation of
motion for this coordinate by integrating out the stable modes, which act as a bath for the quantum mechanical degree of freedom
$q_{-1}$. We will restrict our study to small departures from the critical
radius and obtain the linearized equation of motion for the expectation value of $q_{-1}$. This is consistent with the definition of the growth rate as the
logarithmic derivative of the radius for small departures from the critical value. As is clear from (\ref{unlag}), in the absence of interactions the Hamiltonian for 
the unstable coordinate $q_{-1}$ is that of an {\em inverted
harmonic oscillator} (of negative frequency squared) and small departures
from the unstable equilibrium value $q_{-1} =0$ will grow exponentially
with the growth rate $\Omega$. 

\subsection{The initial value problem:}


We study the time evolution of the expectation value of  the unstable coordinate $Q =\langle q_{-1} \rangle$ by 
proposing an initial state described by a density matrix which is a tensor product of the density matrix for the
unstable coordinate and a density matrix that describes all the stable modes in thermal equilibrium at a given
temperature. Therefore 
\begin{equation}
\rho(0)= \rho_{-1}(0)\otimes \rho_s(0) \label{rhoinitial}
\end{equation}
Where $\rho_{-1}$ describes a pure state in which the expectation value of the unstable coordinate  is $Q_0$, and $\rho_s(0)$ is a thermal density matrix for the stable modes. As mentioned above the translational mode is clamped. 
The time evolution of the initially prepared density matrix is given by Liouville's equation, whose solution is
\begin{equation}
\rho(t)= U(t)\rho(0)U^{-1}(t) \label{timeevoldens}
\end{equation} 
with $U(t)$ being the unitary time evolution operator. As described in detail in reference\cite{boysinghlee} the time
evolution can be cast in terms of a time dependent Hamiltonian in which the interaction is switched on at the initial
time $t=0$. Alternatively the initial value problem can be cast in terms of linear response to an
external source adiabatically turned-on from $t=-\infty$
 that determines the initial preparation\cite{rgir}. Both approaches are equivalent and  the reader is referred to\cite{boysinghlee,rgir} for more details on the initial value problem. 


The equation of motion in real
time is obtained by using the generating functional of non-equilibrium
Green's functions which requires a path integral along a contour in
complex time and the following effective Lagrangian\cite{keldysh,nos}:
\begin{equation}
L_{eff}= L[q^+_{-1},q^+_p]-L[q^-_{-1},q^-_p]
\end{equation} 
where the labels $\pm$ refer to the forward $(+)$ and backward $(-)$ branches
along the complex time contour\cite{keldysh,nos}. The equation of motion
for the expectation value $Q(t) = \left<q_{-1}(t)\right>$ is obtained by performing the shift
\begin{equation}
q^{\pm}_{-1}(t) = Q(t) + \tilde{q}^{\pm}(t) ~~;~~ \left<\tilde{q}^{\pm}(t)\right>=0 \label{shift}
\end{equation}
Imposing the condition $\left<\tilde{q}^{\pm}(t)\right>=0$ to all orders in perturbation
theory leads to the retarded equation of motion for $Q(t)$. For a detailed 
presentation of this method in many other situations see\cite{nos}. 

The important ingredients in this program are the real time Green's functions
for the stable coordinates which are assumed to be in thermal equilibrium. These are the following

\begin{equation}
\begin{array}{lclcl}
\left< q^+_p(t)q^+_{p'}(t^\prime) \right> & =& -i\,\delta_{p,-{p'}} G^{++}_{p}(t,t^\prime) & = &   
-i \delta_{p,-{p'}} \left[{\cal G}^>_p(t,t')\Theta(t-t')+{\cal G}^<_p(t,t')\Theta(t'-t)
\right] \\
& & & & \vspace{-2ex}\\
\left< q^-_p(t)q^-_{p'}(t^\prime) \right> & =& -i\,\delta_{p,-{p'}}G^{--}_{p}(t,t^\prime) & = &  
-i \delta_{p,-{p'}} \left[{\cal G}^>_p(t,t')\Theta(t'-t) + {\cal G}^<_p(t,t')\Theta(t-t')
\right]  \\
& & & & \vspace{-2ex}\\
\left< q^+_p(t)q^-_{p'}(t^\prime) \right> & =& i\,\delta_{p,-{p'}}G^{+-}_{p}(t,t^\prime) & = &  
-i\,\delta_{p,-{p'}}\,{\cal G}^<_p(t,t')  \\
& & & & \vspace{-2ex}\\
\left< q^-_p(t)q^+_{p'}(t^\prime) \right> & =& i\,\delta_{p,-{p'}}G^{-+}_{p}(t,t^\prime) & = & 
-i\, \delta_{p,-{p'}} \, {\cal G}^>_{p}(t,t')= -i\, \delta_{p,-{p'}} \,{\cal G}^<_p(t',t),  
\end{array}
\label{greens}
\end{equation}
where
\begin{eqnarray}
{\cal G}^>_p(t,t') & = & \frac{i}{2\omega_p} \bigg[ \left(1\,+\,n_p\right)\mbox{exp}
\left\{-i\omega_p(t-t')\right\}
\,+\,n_p\,\mbox{exp}\left\{i\omega_p(t-t')\right\} \bigg] \nonumber \\
{\cal G}^<_p(t,t') & = & \frac{i}{2\omega_p} \bigg[ \left(1\,+\,n_p\right)\mbox{exp}
\left\{i\omega_p(t-t')\right\}
\,+\,n_p\,\mbox{exp}\left\{-i\omega_p(t-t')\right\} \bigg]\nonumber \\
n_p & =  & \frac{1}{ \mbox{exp}\left\{ \beta\omega_p \right\}\,-\,1 }. 
\label{ggreater}
\end{eqnarray}

We will carry our derivation of the equation of motion in the linear theory where we will neglect nonlinear terms, ${\cal O}(Q^2)$ and higher, and 
furthermore we will neglect the self-interaction of the unstable coordinate
(cubic and quartic terms). As mentioned before, the linearization of the equation of motion is consistent
with the focus on the dissipative corrections to the growth rate, which
is defined for small departures from the critical radius.

In this case, the non-equilibrium effective action $S_{eff}$ becomes
\begin{eqnarray}
\label{acao1}
S_{eff} &=&\int_{-\infty}^{+\infty} dt \,\left\{L_0[\tilde{q}^+(t)] + L_0[q^+_p(t)]\,+\,
\tilde{q}^+(t) \left[ -\ddot{Q}(t) +(\Omega^2 + \delta\Omega^2)Q(t) - \sqrt{\lambda}~\theta(t) 
\sum_{p,p'}{\cal B}_{p,p'}q_p^+(t)q_{p'}^+(t) \right. \right.\nonumber \\
&& \left. \left.
\,-\, 2\lambda ~ \theta(t) \sum_{p,p'}{\cal A}_{p_1 p_2}
 q_{p_1}^+(t)q_{p_2}^+(t) Q(t) \,+\,
h  \right]
\;-\;\sqrt{\lambda}~\theta(t) \, Q(t) \sum_{p,p'}{\cal B}_{pp'}
q_p^+(t) q_{p'}^+(t) \;-\; \left[ (+)\leftrightarrow (-) \right]\; \right\},
\nonumber \\
& &
\end{eqnarray}
where we have written only the terms that are relevant to the lowest
order calculation ${\cal O}(\lambda)$ which is the focus of the present
discussion and included the time dependence of the Hamiltonian to set up the initial value problem\cite{boysinghlee}. 

Imposing the condition $\langle \tilde q^+_{-1} (t)\rangle =0$, 
up to ${\cal O}(\lambda)$, we obtain the following equation to
one-loop order  
\begin{eqnarray}
&&i \int_{-\infty}^\infty dt'  \left< \tilde q^+ (t) \tilde q^+ (t') \right> \left[\ddot{Q} (t') 
\, - \, \Omega^2 \, Q(t')-  \Delta Q(t')-{\cal H}
-2i\lambda \sum_{p_1,p_2,p_3,p_4}{\cal B}_{p_1 p_2}{\cal B}_{p_3 p_4} \times
\right.  \nonumber \\
&&  \left. \int_{-\infty}^{+\infty}~dt''~\bigg\{ 
\langle q_{p_1}^+(t')q_{p_3}^+(t'')\rangle 
\langle q_{p_2}^+(t')q_{p_4}^+(t'')\rangle 
- \langle q_{p_1}^+(t')q_{p_3}^-(t'')\rangle
\langle q_{p_2}^+(t')q_{p_4}^-(t'')\rangle\bigg\} Q(t'')\right] \; = 0 \nonumber 
\end{eqnarray}
with
\begin{eqnarray}
\Delta & = &\delta \Omega^2 -2\lambda \sum_{p,p'}{\cal A}_{p,p'}<q^+_p(t')q^+_{p'}(t')> \nonumber \\
{\cal H} & = & h - \sqrt{\lambda}\sum_{p,p'}{\cal B}_{p,p'}<q^+_p(t')q^+_{p'}(t')> 
\label{eqnmot}
\end{eqnarray}
Since the correlation $ \left< \tilde q^+ (t) \tilde q^+ (t') \right>$ does not vanish 
for all values of $t$ and $t'$, this implies that the quantity between the square brackets must vanish\cite{nos}. Furthermore, we now choose the counterterm
in such a way to ensure that ${\cal H}=0$. We thus obtain the  equation 
of motion for the expectation value of the unstable coordinate for
$t>0$

\begin{equation}
\ddot{Q} (t) \,-\, \Omega^2 \, Q(t)-\Delta Q(t) -
\lambda \, \int_{0}^{t}~dt'~ \Sigma (t-t') Q(t')=0 ~~ ; ~~
\dot{Q}(t=0) =0 ; Q(t=0)=Q_0
\label{eqmv2} 
\end{equation}
where the self energy $\Sigma (t-t')$ is given by
\begin{eqnarray}
\Sigma (t-t')  
& = & \sum_{p,p'} \frac{{\vert{\cal B}_{pp'}\vert}^2}{\omega_{p}\omega_{p'}}
\left\{(1+n_p+n_{p'}) \sin[(\omega_p+\omega_{p'})(t-t')] \,-\,
(n_p-n_{p'}) \sin[(\omega_p-\omega_{p'})(t-t')] \right\},\label{sigmat} \end{eqnarray}
and we have used the fact that $ {\cal B}^*_{p,p'}\,=\,{\cal B}_{-p,-p'}$.
The two different terms in the self energy, proportional to the sum and difference of frequencies
respectively have a simple but important interpretation. The first term proportional to the sum of
frequencies corresponds to the process in which the coordinate $q_{-1}$ ``decays'' into two mesons with probability $(1+n_p)(1+n_{p'})$ minus the ``recombination'' process with probability
$n_p n_{p'}$. The second term proportional to the difference of frequencies originates in Landau damping
and corresponds to the scattering of the unstable coordinate with mesons in
the medium, with probability $(1+n_{p'})n_p$ minus the reverse process with probability $(1+n_{p})n_{p'}$\cite{weldon}.  


The counterterm $\delta \Omega^2$ will be chosen to cancel the tadpole 
contribution to $\Delta$ (see eqn. (\ref{eqnmot})) 
 as well as the {\em static} contribution from the self-energy, since
this contribution is a {\em static} renormalization of the unstable frequency associated with a stationary bubble.   


\subsection{Viscosity corrections to the  growth rate}

In order to obtain the influence of thermal and quantum fluctuations on 
the growth rate of the bubble, we must solve the equation of motion 
(\ref{eqmv2}). This is achieved via the Laplace transform. Introducing
the Laplace transforms of $Q(t), \Sigma(t)$ as
$\tilde{Q}(s), \tilde{\Sigma}(s)$ respectively with
$s$ the Laplace variable. The Laplace transform of (\ref{eqmv2})  with the specified initial conditions is given by 

\begin{equation}
\tilde{Q}(s)=\frac{s \, Q_0}
{s^2-(\Omega^2+\Delta) -\lambda \tilde{\Sigma}(s)},
\label{eqsI}
\end{equation}
where $\tilde{\Sigma}(s)$ is given by

\begin{equation}
\tilde\Sigma(s) = \sum_{p,p'}\frac{|{\cal B}_{p,p'}|^2}{\omega_{p}\omega_{p'}}
\left\{(1+n_p+n_{p'})\frac{\omega_p+\omega_{p'}}{s^2+(\omega_{p}+\omega_{p'})^2}  \,-\,
(n_p-n_{p'})\frac{\omega_p-\omega_{p'}}{s^2+(\omega_{p}-\omega_{p'})^2} \right\}
\label{sigmas}
\end{equation}


We now isolate the static contribution by substracting 
\begin{equation}
\tilde{\Sigma}(0) = lim_{s\rightarrow 0} \tilde{\Sigma}(s) \label{subs}
\end{equation}
from the Laplace transform of the self-energy. Using the identity
\begin{equation}
lim_{s\rightarrow 0} \frac{W}{s^2+W^2} = {\cal P}\left(\frac{1}{W}\right) \label{PP}
\end{equation}
we now fix the counterterm $\delta \Omega^2$ such that $\Delta + \lambda \tilde{\Sigma}(0)=0$. Introducing  
\begin{equation}
\tilde\Gamma(s) = \sum_{p,p'}\frac{|{\cal B}_{p,p'}|^2}{\omega_{p}\omega_{p'}}
\left\{  
\frac{1+n_p+n_{p'}}{\omega_p+\omega_{p'}}
\frac{s^2}{s^2+(\omega_{p}+\omega_{p'})^2}  \,-\,
(n_p-n_{p'}){\cal P} \left(\frac{1}{\omega_p-\omega_{p'}}\right)
\frac{s^2}{s^2+(\omega_{p}-\omega_{p'})^2} \right\}
\label{gamas} 
\end{equation} 
the Laplace transform of the equation of motion becomes
\begin{equation}
\tilde{Q}(s)=\frac{s\,Q_0}
{s^2-\Omega^2 +\lambda \tilde{\Gamma}(s)},
\label{eqs}
\end{equation}

In order to avoid cluttering of notation we will not write explicitly the ${\cal P}$ in eqn. (\ref{gamas}) in what follows, 
 but it must always
be understood that the term $1/(\omega_p-\omega_{p'})$ actually refers to its principal part. This principal part
prescription arises from the subtraction in the limit of vanishing $s$ which is the equivalent of the frequency in the
real time domain, since the Laplace transform requires that $s\rightarrow i\omega +\epsilon$ with $\omega$ the frequency\cite{boysinghlee}. 

At this stage it becomes clear that the procedure of absorbing the local (zero frequency limit) contributions to the self
energy in the {\em static} renormalization of the growth rate $\Omega$ provides the correct description of the
dynamics. Viscosity and dissipative effects only arise from the time dependence of the expectation value of
the coordinate and the {\em static} medium renormalization had already been absorbed into the  definition of 
the growth rate as the limit of zero frequency. 
Viscosity and dissipation arise from the transfer of energy of the bubble wall to other excitations
(mesons) and the growth slows down because of these processes. An important distinction arises in this case as compared
to the familiar situation in field theory in which complex poles determine the renormalization to the mass from the
{\em real} part of the self-energy on shell and the decay width from the {\em imaginary} part of the self-energy on shell.
In this case, however, because the frequency associated with the unstable mode is {\em imaginary}, the real part of
the self-energy will renormalize the {\em growth rate} whereas the imaginary part (if any) at the position of the 
(purely imaginary pole) will provide {\em oscillatory} contributions to the bubble dynamics.  Therefore, we emphasize
that {\em viscosity} effects that will diminish the growth rate are determined by the {\em real part of the self-energy}
at the position of the pole. This is a striking difference from the usual case in which damping and viscosity are
associated with the imaginary part of the self-energy on shell.


The real time dynamics of the bubble growth, $Q(t)$,  is given by the inverse Laplace transform 
\begin{equation}
Q(t) = \frac{1}{2\pi i} \int_C e^{st} \; \tilde{Q}(s)  \;ds \;,
\label{contint}
\end{equation}
where $\tilde{Q}(s)$ is given by eq.(\ref{eqs}) and $C$ refers to the Bromwich 
contour running along the imaginary
axis to the right of all the singularities of $\tilde{Q}(s)$ in the complex 
$s$-plane.  

The analytic structure of $\tilde{Q}(s)$ consists of cuts along the imaginary 
axis in the $s$-plane and  poles. The two different processes of decay into meson pairs (and recombination)
and Landau damping discussed above yield  
two different cut structures. 

The first term in eq. (\ref{gamas}) gives a two-meson cut and the contribution from Landau damping determines a cut structure that includes the origin in the s-plane. The structure of these cuts depend on
the matrix elements of the interaction as well as the full spectrum of excitations and will be investigated in
detail for particular cases in the next sections. 

The growth rate of the bubble with the quantum fluctuation effects included is given by 
the pole $s_p$ of $\tilde{Q}(s)$, eq. (\ref{eqs}), which satisfies the following relation
\begin{equation}
s_p^2\,-\,\Omega^2\,+\,\lambda  \tilde{\Gamma}(s_p) \,=\, 0. \label{polecondition}
\end{equation}

To first order in $\lambda$, the pole of $\tilde{Q}(s)$ which corresponds to 
a growing bubble is given to lowest order (${\cal O}(\lambda)$) by
\begin{equation}
s_p = \pm(\Omega +\lambda\Omega^{(1)})
\end{equation}
where the first order quantum and thermal fluctuation correction $\lambda\Omega^{(1)}$ to the bubble growth is 
given by
\begin{eqnarray}
\lambda \Omega^{(1)} &= & -\lambda  \frac{\tilde{\Gamma}(\Omega)}{2\Omega} \nonumber \\
&=& - \frac{\lambda}{2}
\sum_{p,p'}\frac{|{\cal B}_{p,p'}|^2}{\omega_{p}\omega_{p'}}
\left\{\frac{(1+n_p+n_{p'})}{(\omega_p+\omega_{p'})}\left( \frac{\Omega}{\Omega^2+(\omega_{p}+
\omega_{p'})^2}  \right) \right. \nonumber \\
& & \hspace{1.35in} - \left.\;\frac{n_p-n_{p'}}{\omega_{p}
- \omega_{p'}}\left(\frac{\Omega}{\Omega^2+(\omega_{p}- \omega_{p'})^2}  \right)
\right\}
\label{delomega}
\end{eqnarray}

We note that both terms inside bracket in (\ref{delomega}) are {\em positive} since $n_p < n_{p'}$ for
$\omega_p > \omega_{p'}$, we recall again that the term $1/({\omega_{p}
- \omega_{p'}})\equiv {\cal P}(1/({\omega_{p}
- \omega_{p'}}))$ from the discussion following eqn. (\ref{gamas}). 
Therefore we conclude that the correction to the growth rate is {\em negative}
i.e. the dissipative effects of the coupling to mesons results in a smaller growth rate of supercritical
bubbles. 


Neither the frequencies of oscillations around the stationary bubble nor
the matrix elements are quantum mechanical in origin. However the one loop
contribution ${\cal O}(\lambda)$ to the self-energy is of quantum origin. 

If we restore $\hbar$ in the expression above this results in
$\lambda \rightarrow \lambda \hbar \; ; \; T \rightarrow T/\hbar$.  
Obviously neither the eigenvalues of the fluctuation operator nor the matrix elements depend on $\hbar$ but the one
loop contribution to the self-energy includes the $\hbar$ from the coupling (one loop) as well as from the temperature
factors. 


\subsection{Classical Limit}

In the next sections we  will see that the correction to the growth rate is dominated by low lying excitations 
and for high temperatures these will be such that  $\omega_p << T$. In this case  we can invoke the classical limit
which is best understood by restoring  $\hbar$ as mentioned above
\begin{equation}
n_k=\frac{1}{e^{\hbar \beta\omega_k}-1} \approx 
\frac{T}{ \hbar \omega_k } >> 1
\end{equation}
In this form, 
\begin{equation}
n_k-n_{k'}\approx \frac{T}{\hbar} \left( \frac{1}{\omega_{k}}-\frac{1}{\omega_{k'}}\right)
=-\frac{T}{\hbar} \frac{\omega_{k}-\omega_{k'}}{\omega_{k}\omega_{k'}}
\end{equation}
and in the high temperature limit we further approximate
\begin{equation}
1+n_k+n_{k'}\approx 1+ \frac{T}{\hbar} \frac{\omega_{k}+\omega_{k'}}{\omega_{k}\omega_{k'}} \approx \frac{T}{\hbar} \frac{\omega_k+\omega_{k'}}{\omega_{k}\omega_{k'}}
\end{equation}
These approximations lead to a simplified expression for the lowest order correction to the growth rate 

\begin{eqnarray}
\delta\Omega 
&=& - \frac{\lambda T}{2 }
\sum_{p,p'}\frac{|{\cal B}_{p,p'}|^2}{\omega_{p}^2 \omega_{p'}^2}
\left\{ \frac{\Omega}{\Omega^2+(\omega_{p}+
\omega_{p'})^2}  + \frac{\Omega}{\Omega^2+(\omega_{p}-
\omega_{p'})^2} \right\}
\label{delomegac}
\end{eqnarray}
where the product $\lambda T$ is independent of $\hbar$ and displays clearly the classical limit.
 
The classical limit will be justified in each particular case in the next sections. The expression (\ref{delomegac}) is the final form of the one loop ${\cal O}(\lambda)$ correction to the growth rate arising from {\em dynamical viscosity}
since all of the static contributions had been absorbed by the counterterms. This is as far as we can pursue in a general
manner without addressing the details of the spectrum of fluctuations around the bubble solution. In the next two sections
we study the details of the model determined by the Lagrangian (\ref{model}) with the
potential (\ref{pot}) for the cases of $1+1$ and $3+1$ dimensions.

\section{The 1+1 Dimensional case}
As mentioned in the introduction, the $1+1$ dimensional case is relevant in statistical and condensed matter physics.
Quantum field theory models based on the Lagrangian (\ref{lagrangeana}) with potentials with a stable and metastable state are proposed to describe the low energy phenomenology of quasi-one dimensional charge density wave systems\cite{yulu},
and therefore their relevance in these physical situations warrants the study of this case.

\begin{figure}
\centerline{ \epsfig{file=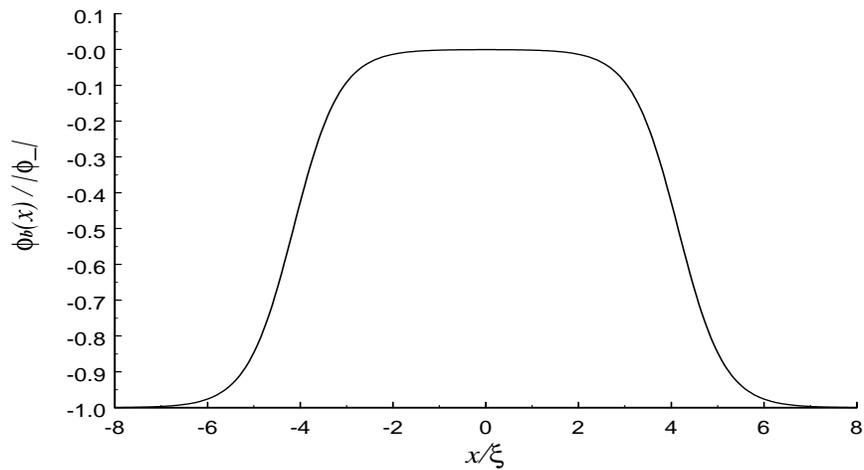,width=4.75in,height=3in}}
\caption{$\frac{\phi_b(x)}{|\phi_-|}$ for a critical bubble in $1+1$ dimensions. $\epsilon=1.001$.\label{critbub}}
\end{figure}

For $V(\phi)$ given by eq. (\ref{pot}), the solution to the static classical equation of motion (\ref{1Dbubble})
for one spatial dimension can be found exactly\cite{boyancondbub,sphal}. 
The critical bubble is found to be given by\cite{boyancondbub,sphal}
\begin{equation}
\label{bubble1d}
\phi_b(x,s_0)=\phi_-+{m\over 2\sqrt{2\lambda}} \left\{\tanh\left[\frac{x}{\xi}
 +s_0\right]-\tanh \left[\frac{x}{\xi}-s_0\right]\right\} ~~;~~ \xi =\frac{2}{m},
\end{equation}
where $\xi$ is the width of the bubble wall and $s_0$ is given in terms of the critical radius $R_c$ by
\begin{eqnarray}
s_0 \,\equiv\, \frac{R_c}{\xi} \,\equiv\, {1\over 2}{\rm cosh^{-1}} \left({\epsilon+1\over\epsilon-1}\right)
\label{s0}
\end{eqnarray}
with $\epsilon$ given by eq. (\ref{epsln}). 
The above solution corresponds to a kink-antikink pair centered at $x = 0$, and separated by 
a distance $R_c$, and is displayed in fig. (\ref{critbub}).
This is the one dimensional bubble that starts at the false vacuum $\phi_-$ almost reaches 
the true vacuum $\phi_+$ and returns to $\phi_-$ and it is similar to the polaron solution
found in quasi-one-dimensional polymers \cite{boyancondbub}.

The total energy of the bubble, given by eq. (\ref{bubenergy}), 
as a function of its dimensionless radius $s$ can be calculated from the field $\phi_b(x,s_0)$ by 
replacing $s_0=R_c/\xi \rightarrow s=R/\xi$ in eq. (\ref{bubble1d}).

The gradient term $(d\phi_b/dx)^2$ contributes to the surface energy
while $V(\phi_b)$ has two contributions; surface contribution and 
volume contribution. Substituting $V(\phi_b)$ in eq. (\ref{bubenergy}) and
evaluating the integral, one finds that the total energy of the bubble is given by
\begin{equation}
E_{var}(s) \,=\, E_{sur}(s)\,+\,E_{vol}(s)
\end{equation}
with
\begin{eqnarray}
E_{sur}(s) & = & {\frac{-4\,{m^3}}{3\,\lambda }} + 
  {\frac{4\,{m^3}\,s }{\left( -1 + {e^{4\,s}} \right) \,\lambda }} + 
  {\frac{3\,{m^3}\,\left( -1 + {e^{8\,s}} - 8\,{e^{4\,s}}\,s \right) \,
      \left( 2 + \eta  \right) }{4\,{{\left( -1 + {e^{4\,s}} \right) }^2}\,
      \lambda }} \nonumber \\
& & \; + \;
  {\frac{4\,{e^{4\,s}}\,{m^3}\,
      \left( 1 - {e^{4\,s}} + 2\,\left( 1 + {e^{4\,s}} \right) \,s \right) }
      {{{\left( -1 + {e^{4\,s}} \right) }^3}\,\lambda }} \\
E_{vol}(s) & = & -{ \frac{\eta \,{m^3} }{\lambda }}\,s
\end{eqnarray}
where $\eta$ is defined as
\begin{equation}
\eta \, \equiv \, \sqrt{\epsilon} \, +\, \frac{1}{\sqrt{\epsilon}} \,-\, 2.
\label{eta}
\end{equation}
The volume contribution grows linearly with the radius of the bubble while
the surface contribution saturates at about the critical radius of the bubble,
$s_0$, and attains the following asymptotic value
\[
\lim_{s \rightarrow \infty} E_{sur}(s) \rightarrow \frac{{m^3}\,\left( 2 + 9\,
\eta  \right) }{12\,\lambda }.
\]
In the thin wall limit ($\eta \rightarrow 0$), the surface energy is simply twice the kink mass\cite{rajaraman},
as the kink-antikink pair becames widely separated  and the exponential interaction between kink and antikink becomes
negligible. 

The total energy of the bubble is depicted in fig.(\ref{e1dfig}). It attains its 
maximum, $E_c$, at the critical radius $s_0$ where  
\begin{equation}
E_c \,=  \frac{m^3}{24\lambda}\left[ 4 + 12\,\eta  + 3\,\eta^2 - 
       3\,s_0\,\eta \,
        \left( 8 + 6\,\eta  + \eta^2 \right)  \right] 
\end{equation}
with
\begin{eqnarray}
s_0 \,=\, {1\over 2}{\rm cosh^{-1}} \left({\epsilon+1\over\epsilon-1}\right)
\,=\, \frac{1}{4}\,\ln\left[\frac{4 + \eta}{\eta}\right]
\end{eqnarray}

In the thin wall limit, the above expressions simplify and we find
\[
\eta \,\simeq \, \frac{(\epsilon-1)^2}{4} \,+\, {\cal O}((\epsilon-1)^3)
\]
and
\begin{equation}
E_c \,= \frac{m^3}{6 \lambda} \,+\, {\cal O}((\epsilon-1)^2)
\end{equation}

 To study the fluctuations around the bubble solution we need 
 the complete set of eigenfunctions $\{{\cal U}_n\}$  satisfying
\begin{equation}
\left[ -\frac{d^2}{dx^2}+V''(\phi_b)\right] {\cal U}_n(x)=\omega_n^2{\cal U}_n(x),
\label{eigen}
\end{equation}
with
\[
V''(\phi_b)\,=\,m^2-\frac{3\,m^2}{2}\mbox{sech}^2\left[\frac{x}{\xi}+s_0\right]
- \frac{3\,m^2}{2}\mbox{sech}^2\left[\frac{x}{\xi}-s_0\right].
\]

Although the spectrum of eigenfunctions and eigenvalues is known exactly in the
case of one kink or antikink\cite{rajaraman}, for the case of the kink-antikink pair
there are not known results that we are aware of. 
Solving for the eigenfunctions $\{{\cal U}_n\}$ in this case is a difficult task but in the 
thin wall limit, the above potential consists of two identical and widely separated wells centered 
at $x=\pm R_c$. The spectrum of each potential well is known in the 
literature\cite{rajaraman}. It consists of a zero frequency mode localized in the well,
 an excited bound state with 
frequency $3m^2/4$ that is also localized in the well and a continuum of scattering states. 

We can use approximate methods such as the linear
combination of atomic orbitals approximation (LCAO) (available in elementary textbooks)
 to provide a reliable estimate for  low lying bound states 
of the above potential from the spectrum of the single potential well. 
In this method, the ground state of the potential $V''(\phi_b)$ is the 
{\em symmetric} linear combination of the ground states of the single well located at 
$x=\pm R_c$ while the first excited state of $V''(\phi_b)$ is the
 {\em anti-symmetric} linear combination.

 In the thin wall limit, these two states are given by
\begin{equation}
{\cal U}_{-1}(x,s_0)=\frac{m}{2\sqrt{E_{c}}}\frac{d\phi_b}
{ds_0}(x,s_0) \;\;\;\; \mbox{with} \;\;\;\; \omega^2_{-1}\,\equiv\, -\Omega^2
\simeq \,-24 m^2 e^{-4 s_0}\,=\,-6 \eta m^2 \label{1dunstmode}
\end{equation}
and
\begin{equation}
{\cal U}_0(x,s_0)=\frac{1}{\sqrt{E_{c}}}\frac{d\phi_b}{dx}(x,s_0)
\;\;\;\; \mbox{with} \;\;\;\; \omega^2_{0}\,= \,0.
\end{equation}
These are the unstable, ${\cal U}_{-1}$, and the zero, ${\cal U}_{0}$, modes which we
discussed earlier, see eqs. (\ref{1dzeromode}) and (\ref{spectrum}). Obviously ${\cal U}_0(x)$ is associated
with translations since it is the spatial derivative of the bubble solution and must correspond to a
vanishing eigenvalue by translational invariance. Since it is antisymmetric there has to be a nodeless eigenfunction of
smaller frequency. The symmetric combination is  ${\cal U}_{-1}(x)$ 
and since the two combinations will be split off in energy by  a
tunneling amplitude that is exponentially small in the
distance between the kink and the antikink, the
unstable frequency must be negative and exponentially small in this separation as
is clearly displayed in eqn. (\ref{1dunstmode}). 

Besides the unstable and the zero modes, there are two bound states that have
energies $\sim (3 m^2/4) \pm \Delta E(s_0)$ with energy difference which is again exponentially small in $s_0$ and correspond to
the symmetric and antisymmetric linear combinations of the bound states of the kink in $\phi^4$ theory\cite{rajaraman}
localized at $x=\pm R_c$.
Since $V''(\phi_b)$ does not exactly reach $m^2$ near the center of the bubble, the potential in the
Schroedinger equation  could allow for shallow bound states near the scattering continuum with binding energies
that are exponentially small in the variable $s_0$. These bound states if
present are extremely difficult to obtain. 

Finally there is the scattering continuum region of the spectrum  characterized by functions ${\cal U}_k$, whose
eigenfunctions are  asymptotically phase shifted plane waves with eigenvalues $\omega_k^2= k^2+m^2$  

In order to compute the matrix elements that enter in the expression for the correction to the growth rate
(\ref{delomegac}) we need either the exact form of the eigenfunctions or an excellent approximation to them. Whereas
we are confident of our analysis regarding the low lying bound states, we lack a full understanding of shallow bound
and continuum states. Such an understanding requires a detailed study of the spectrum which certainly lies beyond the
scope of this article. Although we do not have a complete understanding of the spectrum of eigenfunctions and therefore
we do not have even a good approximation to the matrix elements, we can, however, provide some physically reasonable 
assumptions complemented with dimensional arguments to provide an estimate for the corrections in this case. 

We begin by noting that in the one kink case, the potential that enters in the Schroedinger equation for the
fluctuation is reflectionless\cite{rajaraman} and the scattering states for the one kink (or antikink) case have
a transmission amplitude  which is a pure phase. 

Thus we expect that in  the thin wall limit when the kink-antikink pair separation is much larger than the width of an isolated kink the wave functions of the scattering states will acquire a phase shift that is at
least twice as large as that in the
one kink case and will have a rather smooth dependence on the kink-antikink separation. Because the potentials for each
kink are reflectionless\cite{rajaraman} we {\em assume} that in the thin wall limit the reflection coefficients of each potential well are very small and 
therefore there are no substantial interference effects in the region between the kink and the antikink. Under these assumptions the matrix elements ${\cal B}_{p,p'}$ will be similar to those calculated in reference\cite{alamoudi2} for a single kink case 
and fall off very fast at large momenta justifying the classical limit\cite{alamoudi2}. Hence, under these suitable assumptions the matrix elements are  smooth
functions of $s_0=R_c/\xi$. The contribution from the two-meson cut, i.e. the first term in the bracket in 
(\ref{delomegac}) is proportional to $\Omega$ because the $\Omega$ in the denominator can be neglected in comparison
with the frequencies for the meson states $\omega_p \approx {\cal O}(m)$. Since in $1+1$ dimensions the coupling $\lambda$
has dimensions of $(\mbox{mass})^2$ the correction to the growth rate arising from the two meson cut is
of the form
\begin{equation}
\delta \Omega_{2mes} \approx -\frac{\lambda T}{m^3}\Omega F[s_0]\label{2meson}
\end{equation}
\noindent with $F[s_0]$ a dimensionless slowly varying function of $s_0=R_c/\xi$ which is rather difficult to calculate and 
can only be obtained from a
detailed knowledge of the eigenfunctions. 

The contribution from the Landau
damping cut is more complicated to extract. The second term in the bracket in (\ref{delomegac}) has the form of 
a Lorentzian and since $\Omega$ is exponentially small it is a function that is strongly peaked at $\omega_p = \omega_{p'}$ and the sum (integral)
over $p,p'$ is dominated by a region of width $\Omega$ near $\omega_p=\omega_{p'}$. Assuming that the matrix elements
are smooth functions of momentum, in this one dimensional case the integral over a small region $\omega_p-\omega_{p'} \approx \Omega$ can be done by
taking a narrow Lorentzian and integrating over the relative momentum within this region\cite{note}, leading to a contribution of the form $-(\lambda T/m^2) C[s_0]$ where $C[s_0]$ is a smooth
function of its argument that can only be calculated from a detailed knowledge of the scattering wavefunctions. Hence
\begin{equation}
\delta \Omega_{LD} \approx -(\frac{\lambda T}{m^2})C[s_0]\label{LD}
\end{equation}
and the total shift in the frequency is given by $\delta \Omega = \delta \Omega_{2mes}+\delta \Omega_{LD}$.

Thus we conclude that although we do not have
a complete knowledge of the eigenfunctions and therefore cannot provide a complete calculation of the correction to the
growth rate, suitable assumptions based on the properties of the spectrum of the one kink case combined with dimensional
arguments suggest that  to lowest order in the coupling and in the
thin wall approximation the growth rate of a slightly supercritical bubble is given by

\begin{equation}
\Omega \approx \frac{4\sqrt{6}}{\xi} e^{-\frac{2 R_c}{\xi}}\left[1-\frac{\lambda T\xi^3}{8} 
F[\frac{R_c}{\xi}]\right]- \frac{\lambda T^2 \xi^2}{4}C[\frac{R_c}{\xi}]+\cdots
\label{estimate}
\end{equation}

\noindent with $F[\frac{R_c}{\xi}], C[\frac{R_c}{\xi}]$  slowly varying dimensionless functions of their
argument. Obviously a full calculation of $F[\frac{R_c}{\xi}], C[\frac{R_c}{\xi}]$ requires a detailed understanding of the spectrum, in particular the scattering states.  However it is clear from (\ref{estimate}) that the validity of
a perturbative expansion places a severe constrain on the coupling constant and the value of the temperature, in particular
the second contribution in (\ref{estimate}) arising from Landau damping gives the leading correction in the thin wall
approximation and signals a potential breakdown of perturbation theory in this limit. A more detailed understanding of
this possiblity requires a better knowledge of the scattering matrix elements, this is an extremely difficult problem that depends on the details of the potential and lies outside the  scope of this article.  


\section{The 3+1 Dimensional case}

\subsection{General aspects:}

We now study the $3+1$ dimensional case which is more relevant from the point of view of particle physics, however
before focusing on a particular form of the potential and bubble profile, we can study fundamental model independent properties of the
$3+1$ dimensional case that will determine very robust predictions for the corrections to the bubble growth.

The static bubble configuration $\phi_b(r,R_c)$ is radially symmetric and satisfies the static equation (\ref{3Dbubble}).

To study quantum fluctuations around the critical bubble configuration, 
we need to find the spectrum of the fluctuation operator ${\cal M}$ which  
 in 3 spatial dimensions is given  by
\begin{equation}
{\cal M} \,=\left. \,-\nabla^2 \,+\, \frac{\partial^2 V(\phi)}{\partial\phi^2}\right|_{\phi_b(r,R_c)}
\label{fluctopr}
\end{equation}

Since the critical bubble solution is radially symmetric, we write the eigenfunctions
${\cal U}_{nlm}(r,\theta,\varphi)$ of the differential operator ${\cal M}$ as a product of spherical harmonics $Y_{lm}(\theta,\varphi)$ and radial functions $\psi_{nl}(r)$ that satisfy

\begin{equation}
\left\{-\frac{d^2 }{d r^2} \,-\, \frac{2}{r} \frac{d}{dr} \,+\,
\frac{l(l+1)}{r^2}\,+\, \frac{\partial^2 V(\phi_{b})}{\partial\phi^2}\right\}\psi_{nl}(r)\,=\,
\omega^2_{n} \psi_{nl}(r)
\label{eigen3d}
\end{equation}

Because of the translational invariance of the Lagrangian (\ref{lagrangeana}), there 
is a 3-fold degenerate zero mode given by the eigenfunction 
$\propto \nabla \phi_b(\vec x,R_c) \propto Y_{1;\pm 1,0} d\phi_b / dr$ which
correspond to  translations of the bubble in 3-dimensions with no energy cost. These are the Goldstone modes of the spontaneously broken translational invariance. This can be easily seen by taking the derivative of eq. 
(\ref{3Dbubble}) with respect to $r$ which results in the following equation
\begin{equation}
\left\{-\frac{d^2 }{d r^2} \,-\, \frac{2}{r} \frac{d}{dr} \,+\,
\frac{2}{r^2}\,+\, \frac{\partial^2 V(\phi_{b})}{\partial\phi^2}\right\}\frac{d\phi_b}{
dr}\,=\,0
\label{zeromod}
\end{equation}

In addition to the Goldstone mode, there are other low-lying excitations corresponding to 
 different values of $l\neq 1$. The eigenfunctions of these excitations 
and their eigenvalues can be 
obtained by writing the $l=1$ term in the above equation as
\begin{equation}
\frac{2}{r^2}\,=\, \frac{l(l+1)}{r^2}\,-\,\frac{(l-1)(l+2)}{r^2}
\label{2overr}.
\end{equation}
and we re-write equation (\ref{zeromod}) in the following form
\begin{eqnarray}
&&\left\{-\frac{d^2 }{d r^2} \,-\, \frac{2}{r} \frac{d}{dr} \,+\,
\frac{l(l+1)}{r^2}\,+\, \frac{\partial^2 V(\phi_{b})}{\partial\phi^2}\right\}
\frac{d\phi_b}{dr}\,= \frac{(l-1)(l+2)}{R^2_c}\frac{d\phi_b}{dr}+ \delta V(r)\frac{d\phi_b}{dr} 
\label{lowener} 
\end{eqnarray}
where
\begin{eqnarray}
&& \delta V(r) = (l-1)(l+2)\left[\frac{1}{r^2}-\frac{1}{R^2_c} \right] \label{deltaV}
\end{eqnarray}
Since the function $d\phi_b/ dr$ is strongly localized at $r=R_c$ in the thin wall limit, the second
term on the right hand side is a {\em small localized perturbation}. Therefore the unperturbed 
 lowest lying eigenvalues are given by
\begin{equation}
\omega^2_{0l}\,=\, \frac{(l-1)(l+2)}{R_c^2}.
\label{goldex}
\end{equation}

In the thin wall limit, $\xi/R_c <<1 $ we find that the lowest order correction  (in $\delta V$) to these eigenvalues  is
 of order ${\cal O} (\xi^2/ R_c^{2})$. 
For details see\cite{langer,wallace} and appendix \ref{correction}. This analysis reveals that there is a band of low lying modes with
eigenfunctions 
\begin{equation}
{\cal U}_{0lm}(r,\theta,\varphi) = \sqrt{N_{0l}} Y_{lm}(\theta,\varphi) \frac{d\phi_b(r,R_c)}{dr} \label{lolyin}
\end{equation}
with $N_{0l}$ the normalization constants, corresponding to the 
 $2l+1$-fold degenerate eigenvalues given by (\ref{goldex}). Using
equations (\ref{wsecdv},\ref{omegarelation},\ref{spectrum}) and (\ref{goldex}) with $l=0$ (corresponding
to the unstable mode) we find the normalization to be given by
\begin{equation}
N_{0l} = \frac{1}{R^2_c\sigma} \label{norma}
\end{equation}
in terms of the critical radius and the surface tension. 

We summarize several noteworthy features of these low-lying solutions:
\begin{itemize}
\item {These excitations become Goldstone modes in the limit as
$R_c \rightarrow \infty$, i.e. in the limit in which the radius of curvature of the bubble goes to infinity.
This statement will be understood in detail below in connection with the case of flat interfaces in $3+1$ dimensions.}

\item{The eigenfunction with the lowest eigenvalue, corresponds to $l=0$, i.e. a spherically symmetric 
solution  with a  negative  eigenvalue given by\cite{langer,kapu,wallace}
\begin{equation}
\omega^2_{00}\,\equiv\, -\Omega^2\,=\, -\, \frac{2}{R_c^2}.
\end{equation}
This is the unstable mode which corresponds to a spherically symmetric  expansion or contraction of the bubble and
therefore corresponds to the unstable functional direction. The coordinate associated with this mode is the displacement
from the critical radius.}

\item{The three Goldstone modes ${\cal U}_{0,1,0}~;~{\cal U}_{0,1,\pm 1}$ are
the translation modes.}

\item{The higher energy modes with $l\geq 2$ are excitations on the surface
of the bubble, or surface waves with energies given by
\begin{equation}
\omega^2_{0l} = \frac{(l-1)(l+2)}{R^2_c} = \frac{\Omega^2}{2}(l-1)(l+2) ; ~~ l \geq 2 \label{quasigold} 
\end{equation}
.}

\end{itemize}

 These low lying modes will play a dominant
role and we will refer to them collectively as ${\cal U}_{0lm}$  with eigenvalues given by
(\ref{goldex}) and whose normalized eigenfunctions (\ref{lolyin}) are simple functions of the bubble configuration which
for the potential (\ref{pot}) in the thin wall approximation are given by eqn. (\ref{spcquasi}) below.

That the modes with $l\geq2$ can be identified as wiggles of the bubble surface, or surface waves, 
can be seen from the following expansion\cite{wallace,safran} where as
discussed before  the translational modes is not included because it is
``clamped'' 
 
\begin{eqnarray}
\phi(r,t) & = & \phi_b(r-R_c) + q_{-1}(t) {\cal U}_{000}(r,\theta,\varphi) + \sum_{l\geq 2;m} 
a_{lm}(t) Y_{lm}(\theta,\varphi) \; \left.\frac{d\phi_b}{dr}\right|_{R_c} + \cdots \nonumber \\
 & \simeq & \phi_b(r-R(\theta,\varphi,t)) + q_{-1}(t) {\cal U}_{000}(r,\theta,\varphi) + \cdots\label{surfaceflucs}
\end{eqnarray}
where
\[
R(\theta,\varphi,t) = R_c - \sum_{l\geq 2;m} a_{lm}(t) Y_{lm}(\theta,\varphi)
\]

This is an important identification that we emphasize: {\em this band of
low-lying modes describes fluctuations of the surface of the bubble}. We will argue below that these excitations dominate the infrared behavior of
the viscosity correction and will provide the largest contribution to the
viscosity coefficient.

The low lying spectrum described above is fairly general and only depends on
the existence of a thin wall bubble. Depending on the form of potential $V(\phi)$, there might be other bound states.

An analysis similar to that leading to the band of low-lying excitations
reveals that for the potential (\ref{pot})
there is another band of rotational bound state excitations that starts
near $\omega^2_{10} \simeq 3m^2/4$. In the thin wall limit the radial wave function for the lowest state in this band is given by the bound state of energy $3m^2/4$ of the $\phi^4$ theory in $1+1$ dimensions\cite{rajaraman} which is
also localized at the wall. The eigenfunctions and eigenvalues for this rotational band of excitations are given by (see appendix A)
\begin{equation}
{\cal U}_{1lm}(r,\theta,\varphi) = \sqrt{N_{1l}} Y_{lm}(\theta,\varphi) \psi_1(r)~
; ~~~ \omega^2_{1l}\approx \frac{3m^2}{4}+\frac{l(l+1)}{R^2_c}+{\cal O}(\frac{\xi}{R_c})
 \label{nextband}
\end{equation}
   
\noindent where $\psi_1(r)$ is given by the bound state  of the theory with potential (\ref{pot}) in one spatial dimension\cite{rajaraman},  the eigenfunctions in this band are given in eq.  (\ref{nextbs}) below. 

Finally there is a  continuum of scattering eigenstates with eigenvalues
$\omega_k^2 = k^2 + m^2$. 

As will be discussed later, the contribution to the growth rate from the rotational band  (\ref{nextband}) and of 
the scattering states is  subleading in the thin wall limit. 

The maximum value of angular momentum available for the low-lying part of the spectrum (\ref{goldex}) 
 is limited by the edge of the continuum spectrum or the
presence of higher bound states, hence $l^2_{max}/R^2_c \leq m^2$ or $l^2_{max} \leq (mR_c)^2 = (R_c/\xi)^2$. Therefore
in the thin wall limit the maximum value of the angular momentum $l_{max}>>1$.

\subsection{Planar interfaces, surface waves and (quasi) Goldstone bosons}
The low lying spectrum of eigenfunctions given by (\ref{lolyin}) with eigenvalues (\ref{goldex}) has a simple
physical origin, which can be understood by noticing that in the limit $R_c \rightarrow \infty$ the discrete spectrum
becomes a continuum. 

In the limit when the radius of the critical bubble is very large $R_c \rightarrow \infty$,
the interface between the two phases $\phi_+$ and $\phi_-$ becomes planar and the two phases become degenerate. 

Let us consider a static planar interface configuration corresponding to a domain wall along the z-axis in 
three spatial dimensions. Such a configuration satisfies the
following equation \cite{rajaraman,wallace}
\begin{equation}
- \frac{d^2 \phi_{w}(z)}{dz^2} + \frac{ \partial V(\phi_{w})}{\partial\phi}\,=\,0
\label{walleqn}
\end{equation}
where $\phi_{w}(z)$ satisfies the boundary condition
$\phi_{w}(z) \rightarrow \phi_\pm$ as $z \rightarrow \mp\infty$ and $z$ is the coordinate
perpendicular to the planar interface.

The quantum fluctuations around the classical static wall solution $\phi_{w}(z)$ are 
given by the spectrum of the differential operator
\begin{equation}
{\cal M} \,=\,-\nabla^2 \,+\, \frac{\partial^2 V(\phi_{w}(z))}{\partial\phi^2}
\label{wfluctopr}
\end{equation}

Since the domain wall only depends on the coordinate z,  the differential
operator for the fluctuations is separable in terms of eigenfunctions 
$\psi_{\vec{q}_{\bot}}(z,\vec{x}_{\bot})=e^{i\vec {q}_{\bot}\cdot \vec{x}_{\bot}}\psi_n(z)$, where $\vec{x}_\bot$ denotes the transverse coordinates to $z$, namely $x$ and $y$, and $\vec{q}_{\bot}$ the transverse momentum. The functions $\psi_n(z)$ are so
lution of the
following eigenvalue problem 

 \begin{equation}
\left[-\frac{d^2}{dz^2}+\vec{q}^2_{\bot} \,+\, \frac{\partial^2 V(\phi_{w}(z))}{\partial\phi^2}\right]\psi_n(z) = 
\omega^2_n(\vec{q}_{\bot})\psi_n(z)
\label{flucwall}
\end{equation}

Taking the derivative of eq. (\ref{walleqn}) with respect to $z$, i.e.
\begin{equation}
\left[ - \frac{d^2}{dz^2} + \frac{ \partial^2 V(\phi_{w})}{\partial\phi^2} \right]
\frac{d \phi_{w}}{dz}\,=\,0
\end{equation}
and comparing to eq. (\ref{wfluctopr}) we see that $d\phi_w/dz$ is the zero mode which corresponds to the translational invariance\cite{rajaraman}.

Therefore the eigenfunctions
\begin{equation}
\psi_{\vec{q}_{\bot}} = \mbox{exp}\left(i \mbox{{$\vec{q}_\bot\cdot \vec{x}_\bot$}} \right) \frac{d \phi_{w}}{dz}
\label{capqs}
\end{equation}
 have 
eigenvalues $\vec{q}^2_\bot$. These are Goldstone modes associated with translational invariance and
 represent excitations of the surface
of the planar interface $\phi_w(z)$ since
\begin{eqnarray}
\phi(\vec{r}) & = & \phi_w(z-z_0) +  \sum_{q_\bot} a_{q_\bot} 
\mbox{exp}\left(i \mbox{\boldmath{$q_\bot . x_\bot$}} \right) \frac{d \phi_{w}}{dz}
\nonumber \\
 & \simeq & \phi_w(z-z_\bot(\mbox{\boldmath{$x_\bot$}})) \nonumber 
\end{eqnarray}
where $z_0$ is the position of the planar interface and 
\[
z_\bot(\mbox{\boldmath{$x_\bot$}}) = z_0 - \sum_{q_\bot} a_{q_\bot} 
\mbox{exp}\left(i \mbox{\boldmath{$q_\bot. x_\bot$}} \right)
\]

This is clearly similar to the case of the spherical bubble eq. (\ref{surfaceflucs})
and describes the same physics, i.e. fluctuations of the surface.
In the case of a planar interface these surface waves are also called
capillary waves, and describe the {\em hydrodynamic} modes of long-wavelength fluctuations of interfaces in systems with two degenerate phases separated by an interface\cite{safran}.

In the case of degenerate phases, such as for example a liquid-gas or
an Ising system, these surface waves are Goldstone modes associated with
the breakdown of translational invariance by the presence of the interface.

For a spherical bubble in the thin wall limit these surface waves acquire
a gap given by the inverse radius (proportional to the Gaussian curvature
of the surface). Therefore in  analogy with the case of interfaces for 
degenerate separated phases, we identify these surface fluctuations as
{\em quasi Goldstone modes}. Since, as argued before the maximum
frequency of the surface waves is $< m$ they are {\em classical} in the
high temperature limit $T>>m$. Hence the surface waves are identified
as {\em classical hydrodynamic fluctuations} of the bubble shape and quasi-Goldstone modes in the
thin wall limit. 

We want to emphasize that these low energy excitations are a robust
feature of the thin wall approximation and are {\em model independent}.  
Having studied in detail the {\em general} aspects of fluctuations around
a thin wall bubble, we now focus on the specific details of the
theory with a potential given by (\ref{pot}) so as to be able
to compute the matrix elements and provide a quantitative analysis of
the viscosity effects.

\subsection{$\phi^4$ theory: specifics}

The critical bubble solution that satisfies eq. (\ref{3Dbubble}) with 
the potential (\ref{pot}) is not in general an elementary function, but 
in the thin wall limit the critical radius of the bubble 
$mR_c \sim R_c/\xi >>1$  and 
the function $d\phi_b / dr$ is localized near $R_c$ which makes the 
``friction" term, $r^{-1} d\phi_b/dr \propto 1/(mR_c) \sim \xi / R_c <<1$.
 In this
limit, the critical bubble solution is found to be
\begin{equation}
\phi_b(r,R_c) \,=\, \phi_-\,+\,{m\over 2\sqrt{2\lambda}} \left\{1 -
\tanh \left[\frac{r-R_c}{\xi}\right]\right\}; ~~~\xi= {2\over m}
\label{bubble}
\end{equation}

It corresponds to a field 
configuration that starts around $r=0$ at the true vacuum $\phi_+$, given by eq. 
(\ref{phipls}), and goes to the false vacuum, $\phi_-$, as $r \rightarrow \infty$ with
a surface width $\xi =m/2$ and a critical radius $R_c$. Having specified the critical bubble 
solution $\phi_b$, we now go back and determine explicitly the general expressions which we 
discussed in the previous sections. 

The total energy of the bubble as a function of its radius $R$, given by eq.(\ref{bubenergy}), 
can be calculated from the field $\phi_b(r,R_c)$ by replacing $R_c$ in eq. (\ref{bubble}) by $R$.
The gradient term $\left(\nabla \phi_{b} \right)^2$ contributes to the surface energy
while $V(\phi_b)$ has two contributions: a surface contribution $V_s(\phi_b)$ and 
volume contribution $V_v(\phi_b)$ given by
\begin{eqnarray}
V_s(\phi_b) & = & \frac{1}{2} \left(\nabla \phi_{b}(r) \right)^2 \,+\,
                  \frac{\eta \,m^4}{32\,\lambda }\,\left\{ \mbox{sech}^2\left[\frac{r-R}{\xi} \right] \left( 3 -
 \tanh\left[\frac{r-R}{\xi} \right] \right) 
\right\} \\
V_v(\phi_b) & = &  -\frac{\eta \,m^4}{8\,\lambda } \left(1 - \tanh\left[\frac{r-R}{\xi} \right] \right)
\end{eqnarray}
where $\eta$ is defined by eq. (\ref{eta}).
Substituting the above expressions in eq. (\ref{bubenergy}) and evaluating the 
integral, we find that the total energy of the bubble is given by
\begin{equation}
E_{var}(R) \,=\,
\frac{ 4 \,\pi}{3} \, V(\phi_+)\,R^3 \,+\,
{\frac{4\pi {R^2}{m^3}}{12\lambda}}\,\left[ 1 + \frac{9\,\eta }{2}
  -  \frac{3\eta }{2}\frac{\xi}{R}+ 
       \left(\frac{\pi^2 - 6}{12}+\frac{3\pi^2\eta}{8} \right)
\frac{\xi^2}{R^2}\right] \label{energy3dbubb}
\end{equation}
where in the thin wall limit
\[
V(\phi_+) \,\simeq \, -\frac{ \eta\,{m^4}}{4\,\lambda }\,+\, {\cal O}((\epsilon-1)^3).
\]

The total energy attains its maximum at
\begin{equation}
R_c \,=\, \frac{1}{12\xi \eta}\left[2 + 9\,\eta   + \sqrt{4 + 36\,\eta  + 45\,\eta^2}\right].
\end{equation}
Using the fact that $\eta$ is small in the thin wall limit, we
 find that the critical radius
$R_c$ is given by 
\begin{equation}
R_c \,=\, \frac{1}{3\,\eta \xi}\,\left[ 1 +  {\cal O}((\epsilon-1)^2)\right]\;\; \rightarrow\;\;
\frac{\xi}{R_c} \,\simeq \,3 \, \eta
\end{equation}
and the total critical energy is given by
\begin{equation}
E_c \,=\,\frac{4 \,\pi m}{81 \,\lambda \, \eta^2}\left(1 \, + \,{\cal O}((\epsilon-1)^2) \right)
\end{equation}
which is equivalent to eqs. (\ref{rc}) and (\ref{ec}) with the surface
tension given by 
\begin{equation}
\sigma \,=\,\frac{m^3}{12 \, \lambda}\label{surfacetension}
\end{equation}

The low-lying fluctuation modes ${\cal U}_{0lm}(\theta,\varphi,r)$ eqn. (\ref{lolyin})  are given by 

\begin{eqnarray}
{\cal U}_{0lm}(r,\theta,\varphi) & = & \frac{\sqrt{6\,m}}{4\,R_c} \,
\mbox{sech}^2\left[\frac{r-R_c}{\xi}\right]
Y_{lm}(\theta,\varphi)
 \;\;\;\mbox{with}\;\;\;\omega^2_{0l}\,=\, \frac{(l-1)(l+2)}{R_c^2}
\left[1+{\cal O} \left(
\frac{\xi^2}{R_c^2}\right) \right]\label{spcquasi}
\end{eqnarray} 

\noindent and the next rotational band of bound states is given by (for details see appendix A)

\begin{eqnarray}
{\cal U}_{1lm}(r,\theta,\varphi) & \simeq & \sqrt{N_{1l}} \,
\mbox{sech}\left[\frac{r-R_c}{\xi}\right]\mbox{tanh}\left[\frac{r-R_c}{\xi}\right]
Y_{lm}(\theta,\varphi) \nonumber \\
& &\mbox{with}\;\;\;\omega^2_{1l}\,=\, \frac{3m^2}{4}+\frac{l(l+1)}{R^2_c}+{\cal O}\left(\frac{\xi}{R_c}\right)
\label{nextbs}
\end{eqnarray}

\subsection{Corrections to the bubble growth rate}

From eq. (\ref{delomegac}), the  corrections from quantum and thermal fluctuations to the bubble growth in the present case has the following form in the classical limit
\begin{equation}
\delta\Omega = -\frac{\lambda T}{2 } \sum_{qlm,q'l'm'}\frac{|{\cal B}_{qlm,q'l'm'}|^2}{\omega_{ql}^2\,\omega_{q'l'}^2}\left\{ 
\frac{\Omega}{\Omega^2+(\omega_{ql}+\omega_{q'l'})^2} 
+\frac{\Omega}{\Omega^2+(\omega_{ql}-\omega_{q'l'})^2} 
\right\}
\label{delomega3d}
\end{equation}
where the index $q$ runs over bound and scattering states and
\begin{equation}
{\cal B}_{qlm,q'l'm'} \, \equiv \, \frac{1}{2\sqrt{\lambda}}\int ~d^3r~
V'''(\phi_{b}){\cal U}_{-1}(r){\cal U}_{qlm}({\mathbf r}){\cal U}_{q'l'm'}({\mathbf r})
\label{bqlm}
\end{equation}
with
\begin{equation}
V'''(\phi_{b})\,=\, - 6 m \sqrt{2 \lambda} 
\tanh\left[\frac{r-R_c}{\xi} \right]
\end{equation}

Since $V'''$ is spherically symmetric, the angular integral leads
to 
\begin{equation}
{\cal B}_{qlm,q',l',m'} = {\cal B}_{q,q',l} \delta_{l,l'}\delta_{m,m'}
\end{equation}
with 
\begin{equation}
{\cal B}_{q,q',l}\, \equiv \, \frac{(3m)^{3/2}}{4\sqrt{\pi}\, R_c} \sqrt{N_{ql}} \sqrt{N_{q'l}} 
\int ~r^2 dr~\tanh\left[\frac{r-R_c}{\xi} \right] 
\mbox{sech}^2 \left[\frac{r-R_c}{\xi} \right]\psi_{ql}(r)\psi_{q'l}(r)
\label{bql}
\end{equation}

We now note that the bound state energies only depend on $l$, and that
the energy of the scattering states is independent of $l$. Therefore
there is {\em no Landau damping contribution} from the bound states as consequence of the principal part prescription
which subtracts the contribution from $\omega_p = \omega_{p'}$ in the Landau damping term as discussed in detail below
equations (\ref{gamas}, \ref{delomega}). Therefore the correction to the growth rate can be written as
$\delta \Omega = \delta \Omega_b + \delta \Omega_s$ with the correction from the bound states given by

\begin{equation}
\delta\Omega_b = -\frac{\lambda T}{2} \sum_{n,n'=0,1} \sum_{l}\frac{(2l+1)|
{\cal B}_{n,n',l}|^2}{\omega_{nl}^2\,\omega_{n'l}^2}
\frac{\Omega}{\Omega^2+(\omega_{nl}+\omega_{n'l})^2} \label{corr3dbound}
\end{equation}

\noindent and that from the scattering states given by

\begin{equation}
\delta\Omega_s = -\frac{\lambda T}{2} \sum_{k,k'\neq 0,1} \sum_{l} \frac{(2l+1)|
{\cal B}_{k,k',l}|^2}{
\omega_{k}^2 \omega_{k'}^2}
\left\{ 
\frac{\Omega}{\Omega^2+(\omega_{k}+\omega_{k'})^2}  \,+\,
\frac{\Omega}{\Omega^2+(\omega_{k}-\omega_{k'})^2}
\right\}
\label{corr3dscat}
\end{equation}

The bound state correction has three contributions; the {\em quasi-Goldstone} modes
contribution corresponding
to surface waves with $\omega^2_{0l} = \Omega^2 (l-1)(l+1)/2; l\geq 2$, the contribution from the
higher energy rotational band of  bound states near $\omega^2 \geq  3m^2/4$ and the mixed contribution
from the quasi-Goldstone and the higher energy bound states modes. The denominators
in $\delta \Omega_b$ are of order $ \Omega^6$ for the quasi-Goldstone modes contribution as compared 
to $m^6$ for the mixed and the higher energy bound states contributions. Since in the thin 
wall approximation $m/\Omega \propto R_c/ \xi >>1$, the largest 
contribution arises from the quasi-Goldstone surface modes with the lowest energy denominators.
 It can be easily seen that the matrix elements 
cannot compensate for the difference in powers of $R_c$ and that in fact for the higher energy bound states these
matrix elements are {\em smaller} than those for the surface modes because the wave functions ${\cal U}_{1lm}$ 
actually vanish at the position of the bubble wall (see eqn. \ref{nextbs})). 

The contribution from the scattering states is also seen to be much smaller than that from the surface waves.
The frequencies for the continuum states $\omega_k \geq m >>\Omega$ and the two meson cut give a contribution of
order $\Omega$ (since in the denominators the $\Omega$ can be neglected as compared to $m$). For the Landau damping
cut, the argument is similar to that of the case of one space dimension. 


For $\Omega <<m$ this contribution has 
a Lorentzian shape of width $\approx \Omega$, and the integral over the momenta can be performed in the narrow
width approximation. The subtraction of the static contribution guarantees that the integral is dominated by the
Lorentzian\cite{note} and the region $\omega_k \approx \omega_{k'}$ can be integrated by changing to relative variables and
now in three spatial dimensions the phase space in the region $\omega_k - \omega_{k'} \approx \Omega$ give extra
powers of $\Omega$ as compared to the one dimensional case. Furthermore, the matrix elements are smooth
functions of the radius of the bubble as can be understood simply by a scattering argument from a sharply peaked
potential in three spatial dimensions. These matrix elements do not introduce any singularity in the limit $R_c/\xi >>1$, hence the contribution of the scattering states is at least proportional to $\Omega$ and is therefore  subleading in the thin wall limit.


This analysis leads to the conclusion that the largest contribution to the correction to the growth
rate is given by the {\em quasi-Goldstone modes}, i.e. the surface waves, since these are the lowest lying
excitations and hence provide the smallest energy denominators. 

 The matrix element ${\cal B}_{0,0,l}$  for the surface waves can be calculated easily and we find
\begin{equation}
{\cal B}_{0,0,l} \,=\,\sqrt{\frac{3 m}{\pi}} \frac{2}{5 R_c^2}
\,=\,\sqrt{\frac{3 m}{\pi}} \frac{\Omega^2}{5}
\end{equation}

\noindent leading to the following correction
\begin{equation}
\delta\Omega_{sw} \,=\, -\frac{6 \lambda \alpha}{25 \pi} \frac{m T}{\Omega}
\end{equation}
where
\begin{equation}
\alpha \,\equiv \, \sum_{l=0}^{R_c/\xi} \frac{2 l + 5}{(l+1)^2 (l+4)^2 \left(1+2(l+1)(l+4) \right)} \simeq 0.039
\label{alpha}
\end{equation}

The above series converges rapidly and only the first few terms  contribute to the 
sum, we have evaluated the sum numerically with $R_c/\xi =10$.

Hence we summarize one of the main results of this article: the lowest order correction to the bubble growth for
three dimensional bubbles is dominated by viscosity effects arising from the excitation of long-wavelength surface
waves and is given by 
\begin{equation}
\delta \Omega = -0.003 \lambda \Omega T\xi \left(\frac{R_c}{\xi}\right)^2 \label{fincorr3d}
\end{equation}

Therefore to lowest order in $\lambda$ (one-loop) and to leading order in the thin wall approximation we find that
the growth rate of slightly supercritical bubbles is given by

\begin{equation}
\Omega = \frac{\sqrt{2}}{R_c}
\left[1- 0.003\lambda T \xi \left(\frac{R_c}{\xi}\right)^2 \right] \label{finalgrowthrate}
\end{equation}

\noindent with $\xi$ and $R_c$ are the width and radius of the critical bubble. This is one of the important results of our study. 
The validity of perturbation theory places a very stringent constraint on the quartic coupling constant in the
thin wall limit $R_c/\xi >>1$ and in the classical limit 
$T/m \sim T\xi >>1$. The validity of the classical limit in this
case is warranted: we are studying the dynamics of nucleation via
thermal activation for temperatures below the critical temperature
$T< T_c \approx m/\sqrt{\lambda}$ but for temperatures much larger than
the energy of the low lying excitations, the relevant regime for
thermal activation is $m/\sqrt{\lambda}> T >>m$ in the weak coupling limit.
 Furthermore
the low-lying excitations with frequencies $<<m$ are obviously classical.

\section{Conclusions and discussion}

The focus of this article is to provide a microscopic calculation of the growth rate of slightly supercritical
nucleation bubbles. The model under consideration is a $\phi^4$ scalar theory with an explicitly symmetry breaking
term that produces a metastable and a stable ground state, we studied the case of
nucleation in $1+1$ dimensions as well as $3+1$ dimensions. The former is relevant in the case of quasi-one dimensional
charge density wave systems and organic conductors. We begin our analysis by obtaining the critical bubble solution by
including finite temperature effects in the potential that enters in the classical equation of motion, counterterms are
added to the Lagrangian to compensate for the finite temperature corrections consistently in a perturbative expansion.  

Our approach to obtaining the growth rate is very different from previous treatments in that we begin by expanding the
quantum field  around the critical bubble in terms of the quadratic fluctuations around the critical bubble configuration.
These fluctuations describe an unstable direction associated with small departures from the critical radius,
translational zero modes and stable fluctuations. The translational modes are anchored by fixing the center of the
bubbles and we treat explicitly the interaction between the coordinate associated with the growth (or collapse) of the bubble with those associated with the stable fluctuations.  We obtain the growth rate by obtaining the effective
linearized equation of motion for the unstable coordinate by integrating out the coordinates associated with the stable fluctuations. Two different approximations are involved; i) a weak coupling expansion in terms of $\lambda$ the quartic self-coupling, 
and ii) the thin wall approximation in terms of $\xi/R_c$ with $\xi$ the width of the bubble wall and $R_c$ the critical radius. The first approximation allows a consistent perturbative expansion of the self-energy of the unstable
coordinate, the second allows a quantitative calculation of the relevant matrix elements, furthermore our analysis reveals
that the important fluctuations are {\em classical} for temperatures $T_c>T>>m$, with $T_c$ the critical temperature and
$m$ the mass of quanta in the metastable phase. 

In the one dimensional case we are able to provide an estimate for the growth rate given by eq. (\ref{estimate}) where
the functions $F[R_c/\xi]~,~ C[R_c/\xi]$ depend in a detailed manner upon the scattering states solutions of the eigenvalue problem
for the quadratic fluctuations, and clearly will depend on the details of the potential. This estimate points out  the potential
breakdown of perturbation theory in the thin wall limit. 

In the case of three dimensions we are able to extract some robust features that trascend the form of the potential
and are solely a consequence of the thin wall limit. In particular we identify a rotational band of low lying excitations
which describe {\em surface waves} i.e. ripples on the surface of the bubble. We establish the connection of these surface
waves to the capillary waves of flat interfaces in the case of degenerate but phase separated  thermodynamic states (such as the Ising or liquid - gas at coexistence), the surface waves are then identified as {\em classical long-wavelength hydrodynamic fluctuations}. The unstable coordinate couples to these hydrodynamic fluctuations and as
a result the friction term arising from the self-energy is dominated by the coupling to these hydrodynamic
modes. Clearly the {\em coupling} of the unstable coordinate to these hydrodynamic modes depends on the model, and in
the case under consideration we find in the thin wall limit and to lowest order in perturbation theory the following expression
for the growth rate
\begin{equation}
\Omega = \frac{\sqrt{2}}{R_c}
\left[1- 0.003\lambda T \xi \left(\frac{R_c}{\xi}\right)^2 \right] 
\end{equation}

We also obtain the effective {\em non-Markovian} Langevin equation for the coordinate describing small departures from the critical radius and establish the generalized fluctuation dissipation relation between the viscosity and the noise kernel.
The noise is correlated on time scales comparable to $\Omega^{-1}$ precisely as a consequence of the coupling to the 
hydrodynamic modes and cannot be treated simply with white (delta function)
correlations. 

{\bf Discussion:} Although we have studied a specific microscopic model, in the case of $3+1$ dimensions we have
been able to identify some robust features that trascend the particular model. These are the existence and dominance
of hydrodynamic fluctuations associated with surface waves in the thin wall limit.
 The coupling of the coordinate associated with small 
departures from the critical radius to these low energy fluctuations  induces friction or viscosity corrections to the
growth rate of slightly supercritical bubbles and in the weak coupling and thin wall limit these fluctuations give the
largest contribution to the friction corrections. 

One of our original motivations is to make contact with previous studies of nucleation as applied to the quark-hadron
phase transition. 
In particular Csernai and Kapusta\cite{kapu} have parametrized the coarse grained free energy that describes a  quark-hadron first order phase transition in terms of a local energy variable that can be identified with our scalar field $\phi$. The
form of the potential taken by these authors coincides with our potential $V(\phi)$ (\ref{pot}), with coefficients
that depend on temperature, just as we have argued in this article. Their form of the critical bubble solution and the
variational energy as a function of the radius of a bubble are very similar to those studied in this article. These authors
have established that for about $1\%$ supercooling the critical radius for such a model is about $R_c \approx 12 \mbox{fm}$
 the width of the wall is about $\xi \approx 0.7 \mbox{fm}$ and the thin wall approximation must be reliable in this
regime.

We can obtain  the quartic self-coupling $\lambda$ in (\ref{pot}) from the parameters used in\cite{kapu} by
relating the width of the bubble $\xi =2/m$ and the surface tension $\sigma$ to $\lambda$ via eqn. (\ref{surfacetension}). 
In\cite{kapu} the value of the surface tension for the particular quark-hadron model is $\sigma = 50 \mbox{MeV}/\mbox{fm}^2$ for $T \approx 200 \mbox{MeV}$, yielding a value $\lambda \approx 6$ which leads to a very large correction to the
growth rate. Certainly the large value of the coupling invalidates the perturbative scheme and we cannot draw a definite
conclusion as to the relevance of our lowest order estimate for the quark-hadron transition beyond the statement that the
viscosity induced by the hydrodynamic fluctuations {\em could} result in a very large (negative) correction to the  growth rate and therefore a rather small nucleation rate.    

 For larger supercooling the critical radius becomes smaller and the thin wall approximation breaks down, but in this
case nucleation and spinodal decomposition will be indistinguishable and homogenous nucleation theory may not be the proper
description. 

However despite the limitations of the perturbative expansion and the thin wall approximation, we have provided
a consistent approach to obtain friction or viscosity corrections to the growth rate from a {\em microscopic}
perspective without invoking a phenomenological description. The observation that in the thin wall limit the most important corrections arise from the coupling to
{\em classical} hydrodynamic fluctuations could perhaps pave the way to a systematic hydrodynamic treatment that would circumvent the weak coupling expansion and allow to extract non-perturbative physics. 

\section{Acknowledgements}  
D. B. thanks the N.S.F for partial support through the grant awards: PHY-9605186 and INT-9512798.
S. M. A. thanks King Fahad University of Petroleum and Minerals (Saudi Arabia) 
for financial support. S. E. J. thanks CNPq for finantial support.  F. I. T. thanks FAPEMIG for  support.
E. S. F. thanks  CNPq and FAPERJ for support and D. G. B. thanks FAPERJ and CNPq for support through a
binational collaboration.

\appendix

\section{Corrections to the quasi-Goldstone modes}
\label{correction}
In this appendix we calculate the corrections to the eigenvalues of the quasi-Goldstone  modes that obey the eq. (\ref{lowener}) using first order perturbation theory. First we write
\begin{equation}
\frac{(l-1)(l+2)}{r^2} =  \frac{(l-1)(l+2)}{R_c^2} + \delta V
\end{equation}
with
\[
\delta V = (l-1)(l+2) \left(\frac{1}{r^2} - \frac{1}{R_c^2} \right)
\]

First order energy correction $E_l^{(1)}$ are given by
\begin{eqnarray}
E_l^{(1)} & = & -\left< {\cal U}_{0lm} | \delta V | {\cal U}_{0lm} \right> \nonumber \\
          & = & -\frac{(l-1)(l+2)}{R_c^2} \int_0^\infty r^2 dr N_{0l} \left(
\frac{d \phi_b}{dr} \right)^2 \left(\frac{R_c^2}{r^2} - 1 \right)
\label{fenecor}
\end{eqnarray}
where we have integrated out the angular degrees of freedom.

For the $\phi^4$ potential given by eq. (\ref{pot}), the normalized wavefunctions are
given by eq. (\ref{spcquasi}) and when it is substituted in eq. (\ref{fenecor}) one 
finds that
\begin{equation}
E_l^{(1)} = \frac{(l-1)(l+2)}{R_c^2} \frac{(\pi^2-6)}{12} \frac{\xi^2}{R_c^2} \;\;\;\; ; 
\;\;\;\; \xi = \frac{2}{m}
\end{equation}
Thus to first order in perturbation theory, the eigenvalues are given by
\[
\omega^2_{0l} = \frac{(l-1)(l+2)}{R_c^2}\left[1+{\cal O} \left(
\frac{\xi^2}{R_c^2}\right) \right].
\] 

A similar treatment can be used for the next rotational band based on the bound state of the
$\phi^4$ theory in one spatial dimension. For this we treat the term
\begin{equation}
\delta V^{(1)} = - \frac{2}{r}\frac{d}{dr} + {l(l+1)}\left[\frac{1}{R^2_c}-\frac{1}{r^2}\right]  
\end{equation}
as a perturbation. The unperturbed wave function is the positive energy bound state of the one
dimensional $\phi^4$ theory\cite{rajaraman} and the first order correction
is obtained as before and is  $l(l+1)/R^2_c+{\cal O}\left(\frac{\xi}{R_c}\right)$.

\section{Langevin equation and fluctuation dissipation relation}

The semiclassical Langevin equation is obtained by performing the path integrals over the
bath degrees of freedom, i.e. the stable modes, thus obtaining a non-equilibrium effective functional for the 
unstable mode. This is achieved consistently in perturbation theory\cite{nos,alamoudi2}, and  to lowest order we find

\begin{equation}
Z[j^+=j^-=0] \,=\, \int {\cal D}q_{-1}^+\,{\cal D}q_{-1}^-\, \;e^{i\int_{-\infty}^{\infty} dt^\prime \left(L_0[q_{-1}^+]\,-\,L_0[q_{-1}^-]\right)} \;{\cal F}[q_{-1}^+,q_{-1}^-],
\label{influence}
\end{equation}
where the Lagrangians $L_0[q_{-1}^\pm]$ are given by eq. (\ref{unlag})
and ${\cal F}[q_{-1}^+,q_{-1}^-]$ is the influence functional\cite{feymanvrnon,caldralgt}  which to one loop order and neglecting
the tadpole contributions is given by 
\begin{eqnarray}
{\cal F}[q_{-1}^+,q_{-1}^-] & = & \mbox{exp}\Bigg\{-i^2 \lambda\, \sum_{p,p'}|{\cal B}_{pp'}|^2 \int 
dt\, dt^\prime  
\biggl[q_{-1}^+(t) \, G_p^{++}(t,t^\prime) G_{p'}^{++}(t,t^\prime)\, q_{-1}^+(t^\prime) \nonumber \\
&  &\,+\;
q_{-1}^-(t) \, G_p^{--}(t,t^\prime)G_{p'}^{--}(t,t^\prime) \, q_{-1}^-(t^\prime) 
 - \;   q_{-1}^+(t) \,G_p^{+-}(t,t^\prime)G^{+-}_{p'}(t,t^\prime) \,q_{-1}^-(t^\prime) 
\nonumber \\
&  &\,-\;
q_{-1}^-(t) \, G_p^{-+}(t,t^\prime) G_{p'}^{-+}(t,t^\prime)\, q_{-1}^+(t^\prime) \biggr] \, \Bigg\}
\end{eqnarray}
where the Green's functions are given by eq. (\ref{greens}).

Introducing the Wigner coordinates or center of mass and relative coordinates, $x(t)$ 
and $r(t)$ respectively given by
\[
x(t)\,=\, \frac{1}{2} \left( q_{-1}^+(t)\,+\,q_{-1}^-(t) \right) \hspace{.3in},\hspace{.3in}
r(t)\,=\,q_{-1}^+(t)\,-\,q_{-1}^-(t).
\]
\noindent the generating functional becomes\cite{caldralgt}-\cite{boydsbfs}
\begin{equation}
Z[0] \,=\, \int {\cal D}x\,{\cal D}r\, \;e^{iS[x,r]}
\label{slanvin}
\end{equation}
with the non-equilibrium effective action given by 
\begin{equation}
S[x,r] \,=\, \int dt \, r(t) \, \left[ - \ddot{x}(t) \,+\, \Omega^2\, x(t)\,+ \lambda 
\int dt^\prime\,\biggl( \Sigma_{\mbox{ret}} (t-t^\prime)\,x(t^\prime) \,+ \frac{i}{2}\,K(t-t^\prime)\,r(t^\prime) 
\biggr) \right]. 
\label{lanaction}
\end{equation}
\noindent where for clarity we have neglected the counterterms as well as the tadpoles arising in the
computation of the influence functional. 
The kernels $\Sigma_{\mbox{ret}}(t-t^\prime)$ and $K(t-t^\prime)$ are given by 
\begin{eqnarray}
\Sigma_{\mbox{ret}}(t-t^\prime) & = & \Sigma(t-t^\prime) \Theta(t-t^\prime)    \nonumber \\
K(t-t^\prime) & = & - \sum_{p_1,p_2} \vert {\cal B}_{p_1 p_2} \vert^2
\left\{   
{\cal G}_{p_1}^>(t,t')  {\cal G}_{p_2}^>(t,t') \,+\,  {\cal G}_{p_1}^<(t,t') 
{\cal G}_{p_2}^<(t,t') \right\} \label{noscor} \\
& = & \sum_{p,p'} \frac{{\vert{\cal B}_{pp'}\vert}^2}{2\omega_{p}\omega_{p'}}
\left\{(1+n_p+n_{p'}+2n_pn_{p'}) \cos[(\omega_p+\omega_{p'})(t-t')] \right.\nonumber \\
&&\;\;\;+\,\left.
(n_p+n_{p'}+2n_pn_{p'}) \cos[(\omega_p-\omega_{p'})(t-t')] \right\},\nonumber
\end{eqnarray}
with $\Sigma(t-t^\prime)$ given by (\ref{sigmat}).

At this stage it proves convenient to introduce the identity 
\[
e^{-\frac{1}{2}\int dt \, dt^\prime r(t) \; K(t-t^\prime)\, r(t^\prime)} \, = \, C(t) \int {\cal D} \xi \;e^{-\frac{1}{2} \int dt \, dt^\prime  \xi(t)K^{-1}(t-t^\prime)\xi(t^\prime) \,+\, i\int dt\,\xi(t) r(t) }
\]
with C(t) being an inessential normalization factor, to cast the 
non-equilibrium effective action of the unstable mode
in terms of a stochastic noise variable with a definite probability
distribution\cite{caldralgt}-\cite{boydsbfs}. Using the above relation, the generating functional becomes
\begin{eqnarray}
Z[0] &=& \int {\cal D}x\,{\cal D}r\,{\cal D} \xi \,P[\xi]\; \mbox{exp} \biggl\{i \int dt\; 
r(t) \biggl[ - \ddot{x}(t) \,+\, \Omega^2 \, x(t)\,+\,\lambda 
\int_{-\infty}^t dt^\prime\, \Sigma (t-t^\prime)\,x(t^\prime)\,+\, \xi(t)
 \biggr] \biggr\}, \nonumber \\
\label{ZLanng}
\end{eqnarray}
where the probability distribution of the stochastic noise, $P[\xi]$, is given by
\begin{equation}
P[\xi]\,=\,\int {\cal D} \xi \;\mbox{exp}\left\{ -\frac{1}{2 \lambda} \int dt \, 
dt^\prime  \xi(t)K^{-1}(t-t^\prime)\xi(t^\prime) \right\}.
\end{equation}
In this approximation we find that the noise-noise correlation function  is given by
\begin{equation}
<< \xi(t) \xi(t^\prime)>>\,=\, K(t-t^\prime),
\end{equation}
which is in general colored, i.e. it is not a delta function $\delta(t-t^\prime)$.

The semiclassical Langevin equation is obtained by extremizing  the effective action  in eq. (\ref{ZLanng}) with respect to $r(t)$
\begin{equation}
\ddot{x}(t) \,-\, \Omega^2 \, x(t)\,-\,\lambda 
\int_{-\infty}^t dt^\prime\, \Sigma (t-t^\prime)\,x(t^\prime)\,=\, \xi(t)
\label{Lang}
\end{equation}
Taking the average of the above equation with the
noise probability distribution $P[\xi]$ and identifying $<<x(t)>>=Q(t)$ yields the equation of motion for
the expectation value of the unstable mode, eq.(\ref{eqmv2}). 

The relationship between the kernels $\Sigma_{\mbox{ret}}(t-t^\prime)$ and $K(t-t^\prime)$ 
constitutes a generalized quantum fluctuation dissipation 
relation. 
This relation is established by considering the time 
Fourier transforms of the functions ${\cal G}_{p}^>(t,t') {\cal G}_{p'}^>(t,t')$ and
${\cal G}_{p}^<(t,t') {\cal G}_{p'}^<(t,t')$ which are denoted by ${\cal G}_{pp'}^>(\omega)$
and ${\cal G}_{pp'}^<(\omega)$ respectively. These Fourier transforms 
obey the KMS condition\cite{fetter}
\begin{equation}
{\cal G}_{pp'}^<(\omega) = e^{-\beta \omega} {\cal G}_{pp'}^>(\omega). 
\end{equation}

Using the above relation we find that $\Sigma_{\mbox{ret}}(\omega)$, the Fourier transform in 
time of the retarded self energy $\Sigma_{\mbox{ret}}(t-t^\prime)$ is given by
\begin{equation}
\Sigma_{\mbox{ret}}(\omega) = 2 \sum_{p,p'} \vert {\cal B}_{p p'} \vert^2 \int \frac{d\omega'
}{2\pi} \frac{ {\cal G}_{pp'}^>(\omega')\left[1-e^{-\beta \omega'}\right]}{\omega-\omega'+i\epsilon},
\end{equation}
leading to the imaginary part
\begin{equation}
Im[\Sigma_{\mbox{ret}}(\omega)] = -  \left(1-e^{-\beta \omega}\right) \sum_{p,p'} 
\vert {\cal B}_{p p'} \vert^2 {\cal G}_{pp'}^>(\omega).
\end{equation}

On the other hand the kernel that determines the noise-noise correlation
function $K(t-t')$ has a Fourier transform given by $k(\omega)$ with
\begin{eqnarray}
k(\omega) & = &  \left(1+e^{-\beta \omega}\right) \sum_{p,p'} \vert {\cal B}_{p p'} 
\vert^2 {\cal G}_{pp'}^>(\omega)\nonumber\\
& = & - \coth\left[\frac{\beta \omega}{2}\right] Im[\Sigma_{\mbox{ret}}(\omega)].
\label{flucdiss}
\end{eqnarray}
The above relation between the Fourier transform of the noise-noise correlation
function and the imaginary part of the self-energy is the generalized fluctuation-dissipation relation.

In particular the contribution from the surface waves to the kernel $K(t-t')$ is given by
\begin{equation}
K_{sw}(\tau) = \frac{12mT^2}{25\pi}\sum_{l\geq 2}^{R_c/\xi} \frac{(2l+1)}{(l-1)^2(l+2)^2} \left\{ 1 + \cos\left[2\sqrt{(l-1)(l+2)}\frac{\tau}{R_c}\right] \right\}
\end{equation}

This function oscillates with constant amplitude on a time scale $\sim R_c$ which is the same time scale for growth
of a supercritical bubble. Therefore the noise term {\em cannot} be taken to be uncorrelated over the time scales
associated with the growth of a bubble, i.e. the noise is {\em colored} and a Langevin description based on a white
noise would miss the long-time correlations. The contribution of the scattering states {\em could} lead to a short
range part of the kernel, but the long time behavior of the kernel will be dominated by the low energy surface waves.

The reason that a Markovian Langevin equation with  white noise  fails to describe the dynamics of
the unstable coordinate is that there are slow hydrodynamic fluctuations with time scales comparable to the growth
rate that couple to the unstable coordinate, precisely the same type of fluctuations that dominate the viscosity or
friction.



%

%
%


\end{document}